\documentclass[twocolumn]{aastex6}

\usepackage{graphicx}
\usepackage{hyperref}
\usepackage{amssymb,amsmath}

\def\lea{\mathrel{<\kern-1.0em\lower0.9ex\hbox{$\sim$}}}
\def\gea{\mathrel{>\kern-1.0em\lower0.9ex\hbox{$\sim$}}}
\newcommand{\lta}{{\>\rlap{\raise2pt\hbox{$<$}}\lower3pt\hbox{$\sim$}\>}}
\newcommand{\gta}{{\>\rlap{\raise2pt\hbox{$>$}}\lower3pt\hbox{$\sim$}\>}}

\begin{document}

\title{First Detection of a Pulsar Bow Shock Nebula in Far-UV: PSR J0437$-$4715$^\dagger$}

\author{Blagoy Rangelov\altaffilmark{1}, 
George G. Pavlov\altaffilmark{2}, Oleg Kargaltsev\altaffilmark{1}, Martin Durant\altaffilmark{3}, Andrei M.\ Bykov\altaffilmark{4, 5, 6}, and Alexandre Krassilchtchikov\altaffilmark{4}}
\altaffiltext{1}{Department of Physics, The George Washington University, 725 21st St, NW, Washington, DC 20052}
\altaffiltext{2}{Pennsylvania State University, 525 Davey Lab., University Park, PA 16802}
\altaffiltext{3}{Sunnybrook Health Sciences Centre, Toronto, Canada}
\altaffiltext{4}{Ioffe Physico-Technical Institute, St.-Petersburg, Russia}
\altaffiltext{5}{Saint-Petersburg State Polytechnical University, St.-Petersburg, Russia}
\altaffiltext{6}{International Space Science Institute, Bern, Switzerland}

\thanks{$^\dagger$ Based on observations made with the NASA/ESA Hubble Space Telescope, obtained at the Space Telescope Science Institute, which is operated by the Association of Universities for Research in Astronomy, Inc., under NASA contract NAS 5-26555. These observations are associated with programs GO 12917 and GO 10568.}

\email{rangelov13@gwu.edu}

\slugcomment{The Astrophysical Journal, to be submitted}
\shorttitle{FUV Bow Shock of PSR\,J0437$-$4715}
\shortauthors{Rangelov et al. 2016}

\begin{abstract}

Pulsars traveling at supersonic speeds are often accompanied by cometary bow shocks seen in H$\alpha$. We report on the first detection of a pulsar bow shock in the far-ultraviolet (FUV). We detected it in FUV images of the nearest millisecond pulsar J0437$-$4715 obtained with the {\sl Hubble Space Telescope}. The images  reveal a bow-like structure positionally coincident with part of the previously detected H$\alpha$ bow shock, with an apex at $10''$ ahead of the moving pulsar. Its FUV luminosity, $L(1250-2000\,{\rm \AA})\approx 5\times 10^{28}$ erg s$^{-1}$, exceeds the H$\alpha$ luminosity from the same area by a factor of 10. The FUV emission could be produced by the shocked ISM matter or, less likely, by relativistic pulsar wind electrons confined by strong magnetic field fluctuations in the bow shock. In addition, in the FUV images we found a puzzling extended ($\simeq 3''$ in size) structure overlapping with the limb of the bow shock. If related to the bow shock, it could be produced by an inhomogeneity in the ambient medium or  an instability in the bow shock. We also report on a previously undetected X-ray emission extending for about $5''$ ahead of the pulsar, possibly a pulsar wind nebula created by shocked pulsar wind, with a luminosity $L(0.5-8\,{\rm keV})\sim 3\times 10^{28}$ erg s$^{-1}$. 

\end{abstract}

\keywords{pulsars: individual (PSR J0437$-$4715) --- shock waves --- ISM: jets and outflows --- ultraviolet: ISM --- X-rays: individual (PWN J0437$-$4715)}

\section{Introduction}

Rotation powered pulsars are known to be sources of magnetized relativistic winds whose interaction with the ambient medium produces spectacular pulsar wind nebulae (PWNe), observable from the radio to TeV $\gamma$-rays \citep{2008AIPC..983..171K,2006ARA&A..44...17G,2013arXiv1305.2552K}. When a pulsar is moving through the interstellar medium (ISM) with a speed exceeding the ISM sound speed, a cometary bow shock is expected to form. In the hydrodynamical approximation (see, e.g.,  \citealt{2001A&A...375.1032B}) the pulsar wind is confined to the interior of the contact discontinuity surface, which separates the shocked ISM from the shocked pulsar wind, while the shocked ISM is confined between the forward shock and contact discontinuity surfaces. The highly relativistic pulsar wind bulk flow experiences a termination shock, where the flow speed drops below the sound speed in the outflow. In the idealized case of initially isotropic wind the termination shock has a bullet shape (see, e.g., Figure~9 in \citealt{2004ApJ...616..383G}, and Figures 1 and 2 in \citealt{2005A&A...434..189B}). In reality,  pulsar winds can be highly anisotropic, with a polar component along the pulsar spin axis and an equatorial component, as demonstrated by X-ray observations of PWNe created by young pulsars (see \citealt{2008AIPC..983..171K} for a review). Therefore, the shape of the shocks and the nebula appearance may depend on the angle between the velocity vector and the spin axis of the pulsar \citep{2007MNRAS.374..793V}. In addition,  inhomogeneities in the ambient medium  \citep{2007MNRAS.374..793V} and ISM entrainment \citep{2015MNRAS.454.3886M} can affect the nebula shape and properties. 

The ISM matter is compressed and heated while passing through the forward shock,  which can lead to excitation of ISM atoms followed by radiative de-excitation in the shocked ISM. Unless the local ISM ahead of the moving pulsar is strongly ionized, one can expect strong emission in Lyman and Balmer lines caused by collisional excitation of most abundant hydrogen atoms by fast ions and electrons in the shocked gas and by electron transfer from neutral hydrogen atoms  to protons. Since emission in Lyman lines is strongly absorbed in the ISM and the Earth atmosphere, bow shocks can be most easily detected in H$\alpha$, the strongest of the Balmer lines.

To date the bow-shaped H$\alpha$ nebulae have been detected around 9 pulsars (see Table 1 in  \citealt{2014ApJ...784..154B}; BR14 hereafter). In addition, about two dozen pulsars exhibit elongated or tail-like PWN morphologies in X-rays and radio \citep{2008AIPC..983..171K}, commonly attributed to synchrotron radiation from shocked pulsar winds. However, bow shocks, expected for such pulsars, were detected in H$\alpha$ from only a few of them (PSRs B1957+20, J2124$-$3358 and, possibly, B1951+32). A possible explanation for the scarcity of  such ``hybrid'' bow shocks is a high degree of ionization of the ambient medium, which might be caused, in some cases, by pre-ionization of the ISM by  radiation from the pulsar and/or the X-ray PWN. 

One of the first H$\alpha$ pulsar bow shocks was detected by \citet{1993Natur.364..603B}\footnote{The bow shock was actually detected in the R band that includes the H$\alpha$ wavelength, $\lambda_{{\rm H}\alpha}= 6563$\,\AA.} ahead of the millisecond (recycled) pulsar J0437--4715 (J0437 hereafter). J0437 is a $5.8$\,ms pulsar with spindown energy loss rate $\dot{E}=2.9\times10^{33} I_{45}$\,erg\,s$^{-1}$ (corrected for Shklovskii effect; $I_{45}$ is the neutron star moment of inertia in units of $10^{45}$ g cm$^2$). The pulsar is in a wide binary system ($P_{\rm bin}=5.74$ d, companion separation $a=1.1\times 10^{12}$ cm) with a nearly circular orbit inclined by an angle of $138^\circ$ to the sky plane. The binary companion is a cool ($T\approx4000$ K) white dwarf (WD) (see \citealt{2012ApJ...746....6D} and references therein). The pulsar and WD masses are $M_{\rm PSR}=(1.44\pm 0.07)M_\odot$ and $M_{\rm WD}=(0.224\pm 0.007) M_\odot$ \citep{2016MNRAS.455.1751R}. J0437 is the closest known pulsar, $d=156.79\pm0.25$\,pc, with an accurately measured proper motion,  $\mu_\alpha\cos\delta =121.439\pm0.002$\,mas\,yr$^{-1}$, $\mu_\delta =-71.475\pm0.002$\,mas\,yr$^{-1}$, corresponding to the transverse velocity $v_\perp=104.14\pm0.17$\,km\,s$^{-1}$ \citep{2016MNRAS.455.1751R}. \citet{2012ApJ...746....6D} reported the results of {\sl Hubble Space Telescope} ({\sl HST}) observations of the pulsar in far-UV (FUV) and near-UV (NUV), and analyzed the pulsar plus WD spectrum  from mid-infrared to X-rays and $\gamma$-rays.

A sharp image of the J0437 H$\alpha$ bow shock was obtained by Andrew Fruchter in 1995\footnote{The image with some comments can be found at \url{http://www.stsci.edu/~fruchter/nebula/}.} with the CTIO 0.9 m telescope (A.\ Fruchter, priv.\ comm.). The image clearly shows a symmetric bow shock structure with the standoff distance of about $10''$ ($2.3\times 10^{16}$\,cm) between the J0437 binary and the leading edge (apex) of the bow shock. \cite{1995ApJ...440L..81B} observed the H$\alpha$ bow shock with the 2.3\,m ANU telescope. They measured the H$\alpha$ photon flux, ${\cal F}_{{\rm H}\alpha} = 2.5\times10^{-3}$\,cm$^{-2}$\,s$^{-1}$, and estimated the pre-shock number density, $n_0\sim 0.2$\,cm$^{-3}$. BR14 observed the J0437 H$\alpha$ bow shock with the Optical Imager on the CTIO 4.2\,m SOAR telescope. They measured a factor of 3--4 higher flux, ${\cal F}_{{\rm H}\alpha} = 6.7\times10^{-3}$\,cm$^{-2}$\,s$^{-1}$ from an apex zone of the bow shock. The flux difference can be partly (but likely not entirely) due to the different bow shock areas used in the analysis. \citet{2002ApJ...569..894Z} searched for an X-ray PWN around J0437 in the {\sl Chandra X-ray Observatory} ({\sl Chandra}) images but did not find it.

In this paper we report on our discovery of a FUV bow shock ahead of J0437. We present an analysis of FUV broadband photometry and prism spectroscopy of the bow shock and place constraints on the shape of its spectrum based on our dedicated \emph{HST} program and archival \emph{HST} data.  We also use archival  {\sl Chandra} data to perform a new search for extended X-ray emission  around J0437.

\begin{figure*}
\centering
\includegraphics[scale=0.332]{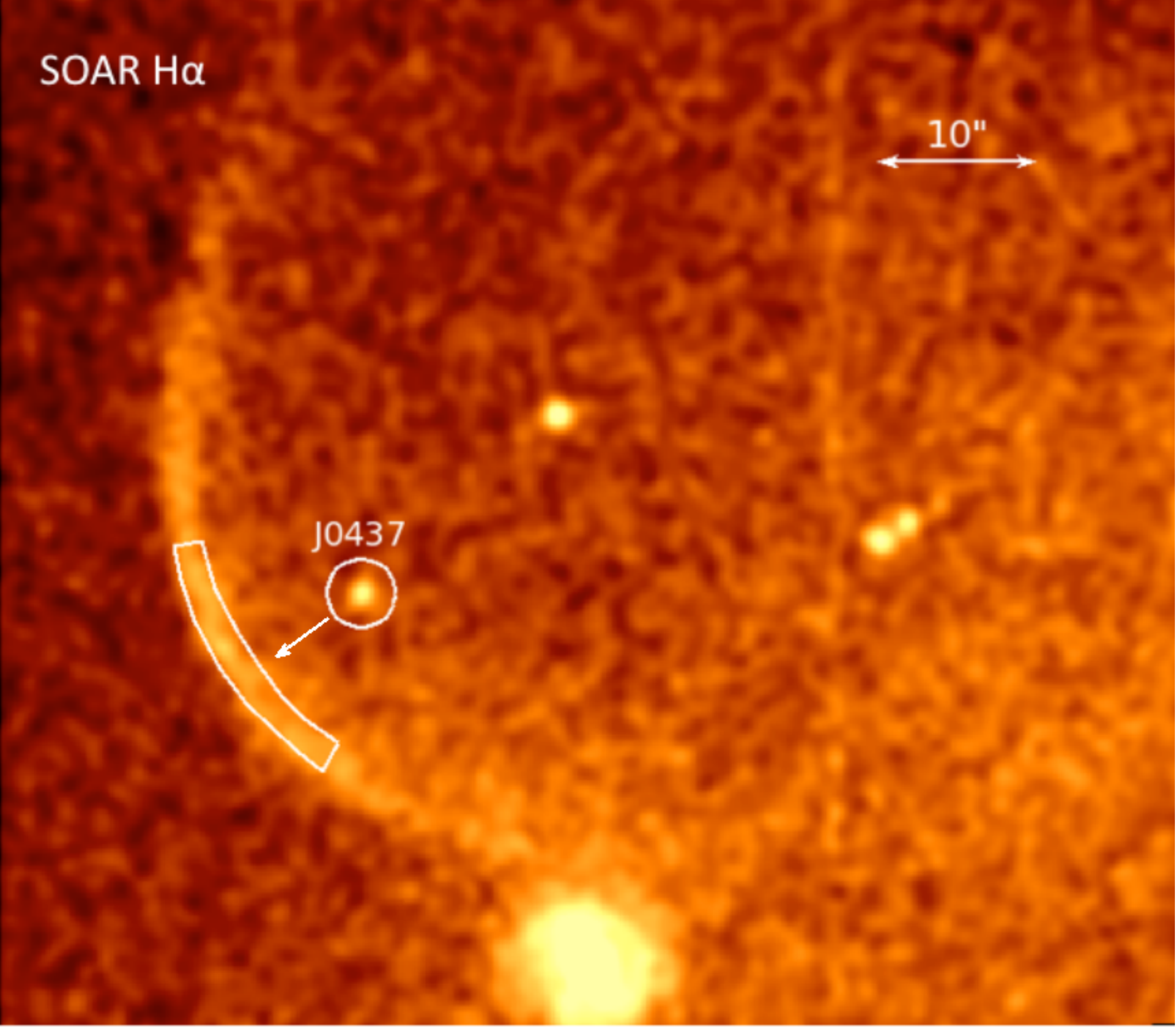}
\includegraphics[scale=0.3]{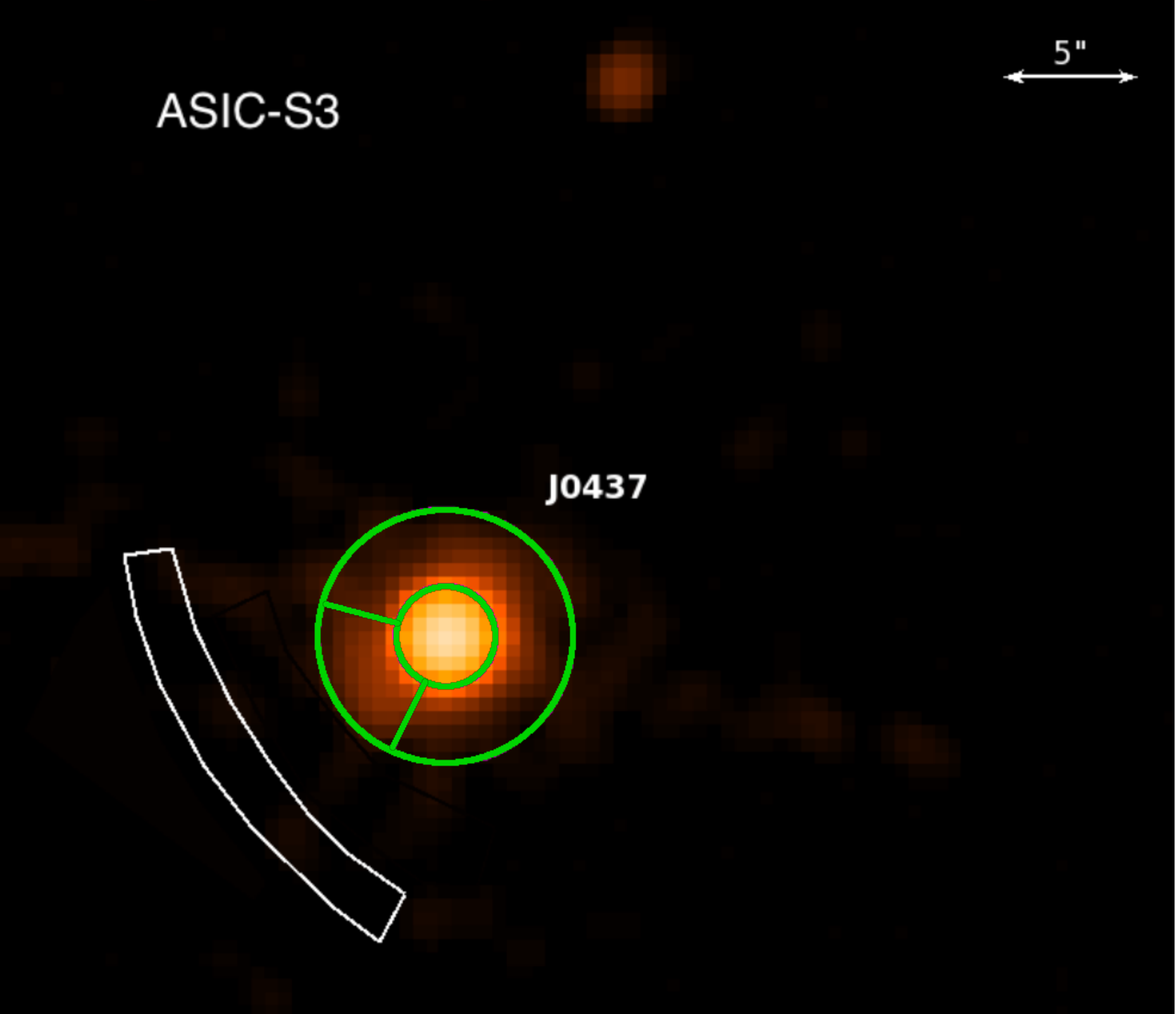}\\
\includegraphics[scale=0.3]{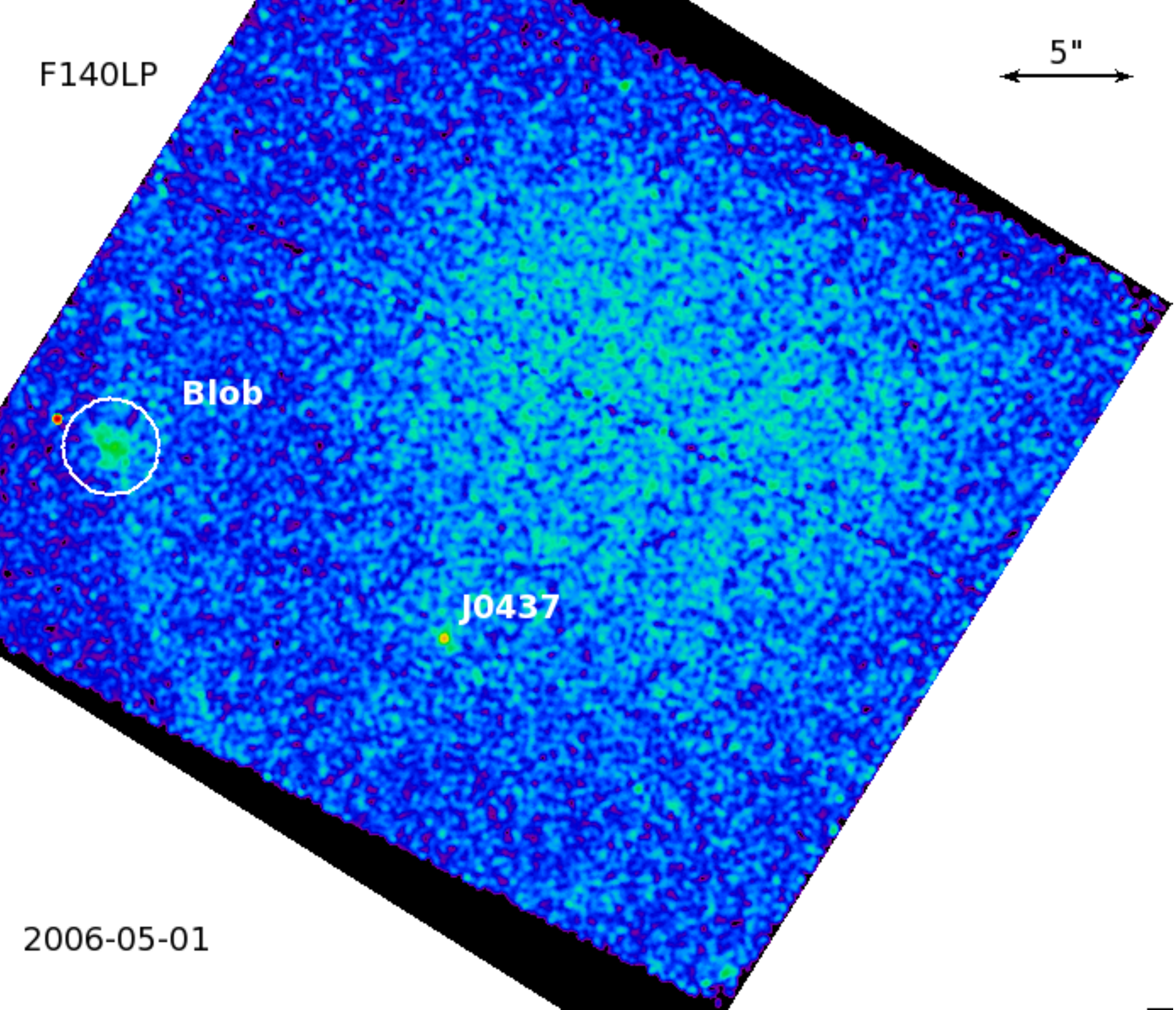}
\includegraphics[scale=0.3]{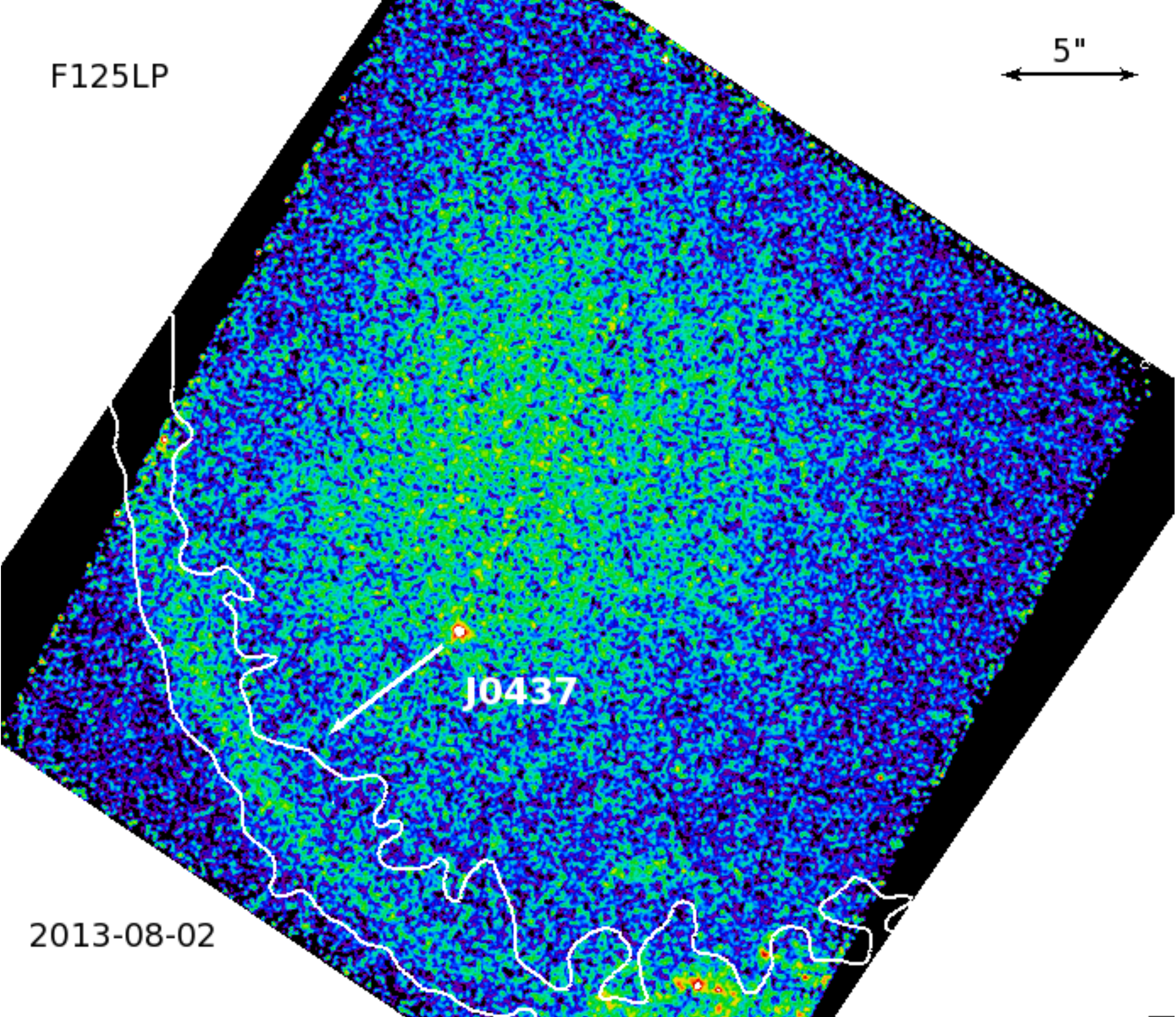}\\
\includegraphics[scale=0.35]{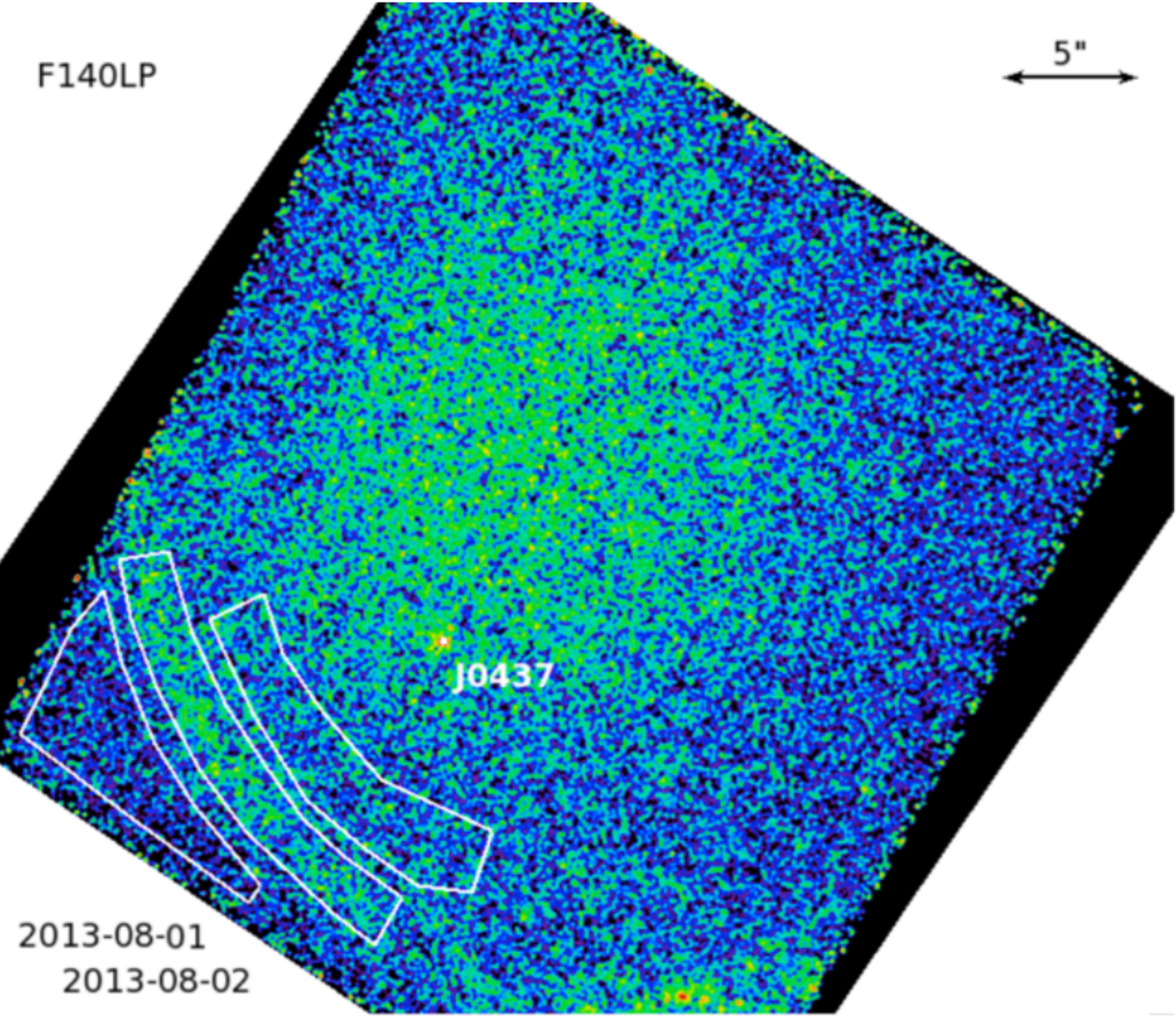}
\includegraphics[scale=0.298]{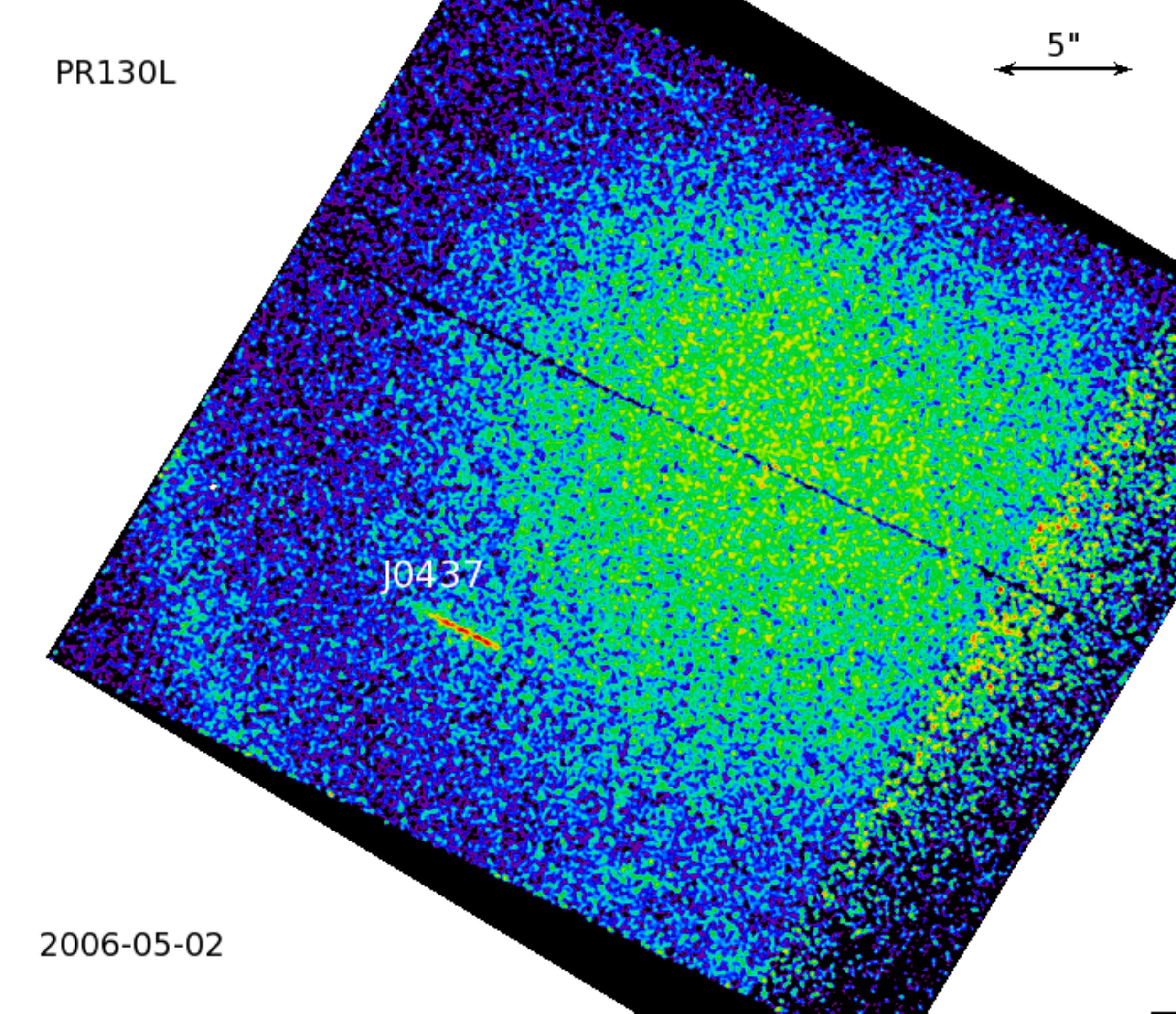}
\caption{Top left panel shows an H$\alpha$ image taken with the SOAR telescope (BR14), smoothed with 5 pixel Gaussian kernel; the region within the white contour was used for the photometric  FUV  bow shock analysis. Top right panel shows a {\sl Chandra} ACIS-S3 image ($0\farcs49$ pixel size,  Gaussian smoothing with $r=1\farcs47$ kernel applied). The green contours delineate the source and background extraction regions used for the extended emission analysis (the source region is the smaller annulus sector around the pulsar, with $r_{\rm in}=1\farcs87$ and $r_{\rm out}=4\farcs68$; the background was measured from the rest of the annulus; see Section 3.2). The FUV extraction region is within the white contour. The other four panels show  FUV images ($0\farcs031$ pixel size, smoothed with $r=0\farcs155$ Gaussian kernel) from different observations and/or exposures. The nonuniform background due to thermal glow is  prominent  in all the four FUV images. The extended ``blob'' (see Section 3.1 for details) is seen in the mid-left panel. H$_{\alpha}$ bow shock contours (based on the SOAR image) are shown in the mid-right panel. The regions used for the FUV flux measurements in  F125LP and F140LP images from the 2013 observations  (see Section 3.3.1) are shown in the bottom left panel. The bottom right panel shows the dispersed PR130L image. All the images, except for the SOAR one, show the same part of the sky, and represent individual observations, with the exception of F140LP from 2013, which are merged images from the two 2013 visits performed within two subsequent days (see Table~\ref{tbl-HST}). North is up, East to the left.}
\label{panel}
\end{figure*}

\section{Observations and Data Reduction}

\subsection{Hubble Space Telescope}

FUV observations of the J0437 field were carried out in two {\sl HST} programs. First program (GO 10568) was devoted to studying the spectrum of the pulsar and WD emission. It involved observations with the High Resolution Camera (HRC), Wide Field Camera (WFC) and Solar Blind Channel (SBC) of the Advanced Camera for Surveys (ACS) with the F140LP (FUV), F300W ($\approx$ U), and F555W ($\approx$ V) filters, and PR130L (FUV) and PR200L (NUV) prisms. The second program (GO 12917) was dedicated to bow shock imaging and photometry  in FUV (F115LP, F125LP, F140LP) and H$\alpha$ (F658N) with the ACS SBC and WFC, respectively. The images from both campaigns were drizzled with the \texttt{AstroDrizzle} package from PyRAF\footnote{See \url{http://www.stsci.edu/institute/software\_hardware/pyraf}.}. Additionally, we used two archival images, F555W ($\approx$ V) and F814W ($\approx$ I), obtained with the Wide Filed Planetary Camera 2 (WFPC2; GO 6642, PI Foster). We downloaded the WFPC2 data from the \emph{Hubble Legacy Archive}\footnote{See \url{http://hla.stsci.edu/}.} (\emph{HLA}). For each filter, the \emph{HLA} combines the individual flat-fielded exposures using the PyRAF  \texttt{Multidrizzle} task, which produces co-aligned, geometrically corrected images. More details about individual observations can be found in Table~\ref{tbl-HST}.

\begin{deluxetable}{llllr}
\scriptsize
\tablecaption{
{\sl HST} observations of the J0437 field.\label{tbl-HST}}
\tablehead{
\colhead{Date} & \colhead{Instrument} & \colhead{PI} & \colhead{Filter} & \colhead{Exp. (s)}}
\startdata
1996-05-19\tablenotemark{$\ddag$} & WFPC2 & Foster & F555W & 560 \\
1996-05-19\tablenotemark{$\ddag$} & WFPC2 & Foster & F814W & 560 \\
2005-12-10\tablenotemark{$\divideontimes$} & ACS/SBC & Kargaltsev & F140LP & 2550 \\
2005-12-10\tablenotemark{$\divideontimes$} & ACS/SBC & Kargaltsev & PR130L & 4800 \\
2006-04-30\tablenotemark{$\ddag$} & ACS/HRC  & Kargaltsev& PR200L & 2637 \\
2006-04-30\tablenotemark{$\ddag$} & ACS/HRC  & Kargaltsev& PR200L & 2396 \\
2006-04-30\tablenotemark{$\ddag$} & ACS/HRC  & Kargaltsev& PR200L & 2580 \\
2006-04-30\tablenotemark{$\ddag$} & ACS/HRC & Kargaltsev & F300W & 360 \\
2006-04-30\tablenotemark{$\ddag$} & ACS/HRC & Kargaltsev & F555W & 200 \\
2006-05-01\tablenotemark{$\intercal$} & ACS/SBC & Kargaltsev & F140LP & 1700 \\
2006-05-01\tablenotemark{$\intercal$} & ACS/SBC & Kargaltsev & F140LP & 2550 \\
2006-05-01\tablenotemark{$\lozenge$} & ACS/SBC & Kargaltsev & PR130L & 4800 \\
2006-05-01\tablenotemark{$\lozenge$} & ACS/SBC & Kargaltsev & PR130L & 3200 \\
2006-05-02\tablenotemark{$\lozenge$} & ACS/SBC & Kargaltsev & PR130L & 1600 \\
2006-05-02\tablenotemark{$\intercal$} & ACS/SBC & Kargaltsev & F140LP & 2550 \\
2006-05-02\tablenotemark{$\lozenge$} & ACS/SBC & Kargaltsev & PR130L & 4800 \\
2006-05-02\tablenotemark{$\intercal$} & ACS/SBC & Kargaltsev & F140LP & 850 \\
2013-01-20\tablenotemark{$\dag$} & ACS/WFC & Durant & F658N &  1997 \\
2013-08-01\tablenotemark{$\ddag$} & ACS/SBC & Durant & F115LP & 4800 \\
2013-08-01\tablenotemark{$\dag$} & ACS/SBC & Durant & F140LP & 2198 \\
2013-08-02\tablenotemark{$\dag$} & ACS/SBC & Durant & F125LP & 4800 \\
2013-08-02\tablenotemark{$\dag$} & ACS/SBC & Durant & F140LP & 2198 \\
\enddata
\tablenotetext{\dag}{Imaging observations analyzed in this paper where the bow shock is detected.}
\tablenotetext{\ddag}{Observations where the bow shock is not detected.}
\tablenotetext{\divideontimes}{Observations where the bow shock strongly overlaps with the thermal glow. These data were not used in our analysis.}
\tablenotetext{\intercal}{The F140LP images where only a small portion of the bow shock is detected at the corner of the CCD. These data were only used in our prism analysis.}
\tablenotetext{\lozenge}{Spectroscopic PR130L observations analysed in Section~3.3.2.}
\end{deluxetable}

The SBC detector, a Multi-Anode Microchannel Array (MAMA) type device sensitive in the 1100\,\AA\ to 2000\,\AA\ range, was the primary instrument in both 2006 and 2013 FUV imaging observations. The details of the spectroscopic 2006 observations and  the spectral analysis of the pulsar itself are given by \cite{2012ApJ...746....6D}. To calibrate the wavelength of dispersed light, direct images (without the dispersing element) with the F140LP filter have been taken in addition to slitless spectroscopic exposures with the PR130L prism. The F140LP ($\lambda\approx1350$--2000\,\AA) filter cuts off all prominent background geocoronal lines. The observations with the PR130L ($\lambda\approx 1230$--2000\,\AA), which provides a modest spectral resolution, were taken in the Earth's shadow to reduce the geocoronal background. The 2006 observations also include HRC NUV imaging and spectroscopy with F300W and PR200L, respectively (see \citealt{2012ApJ...746....6D} for details).

In the 2006 campaign, for the majority of the exposures, the telescope was oriented  to keep the pulsar away  from the region of enhanced SBC thermal glow\footnote{The glow is strongly dependent on the detector temperature, and its contribution to the detector background can range from negligible to  dominating for the central region of the detector.}. As a result, the bow shock (which, at that time, was not expected to be seen in  FUV) was outside the SBC field of view, except for a small part imaged in the detector corner. In one of the visits during the 2006 campaign the observatory experienced loss of guiding stars and simultaneous increase of the detector background. Although most of the data from that visit were lost, the very first F140LP image  of that visit (see Figure~\ref{panel}, middle left panel) happened to be good, with the telescope being oriented in such a way that most of the bow shock was imaged in the region of low detector background.

In the  2013 campaign the  bow shock was intentionally placed outside the region of brightest SBC glow (see Figure~\ref{panel}, middle right and bottom left panels). The  observations with the F115LP (1150--2000\,\AA) and F125LP (1250--2000\,\AA) filters were taken in the Earth shadow to reduce the geocoronal background, while the F140LP filter was used in the non-shadow parts of the respective orbits. A shorter WFC observation used the H${\alpha}$ filter (F658N). Unfortunately, because of a very bright geocoronal  Ly-${\alpha}$  background (despite being in the Earth shadow), we had to discard the F115LP image from further analysis.

\subsection{Chandra X-ray Observatory}

For the J0437 PWN search, we used archival observations with the ACIS detector, which is more sensitive than the HRC detector and provides spectral information. There are 9 ACIS observations of J0437 in the {\sl Chandra} archive. In the earliest observation (ObsID 741, PI Pavlov, the results are reported by \citealt{2002ApJ...569..894Z}) the target was offset by 3\farcm9 from the optical axis to mitigate the pile-up effect, which considerably distorted the Point Spread Function (PSF) and strongly complicated the search for a compact PWN. The other 8 observations were carried out by the {\sl Chandra} ACIS instrument team for calibration purposes. Four of those observations were taken at roll angles too close to $215^\circ$ or $35^\circ$, at which the so-called ACIS trailed image (also known as ``readout streak'')\footnote{See \url{http://cxc.harvard.edu/proposer/POG/html/chap6.html\#tth\_sEc6.12.1}.} of the bright pulsar coincides with the pulsar's sky trajectory and can mimic the expected PWN tail behind the pulsar and/or contribute to the PWN head image. Among the remaining four, one observation had a very large, nonstandard frame time of 6 s (hence the pulsar image was strongly distorted by pileup) and another one had a short exposure of 7.8 ks. Therefore, we selected  two of the calibration observations, ObsIDs 6154 and 6155 (22.8 and 24.8 ks exposures, $283^\circ$ and $271^\circ$ roll angles, respectively) for the PWN search. The data were taken on 2005 February 18 (ObsID 6154) and 2005 March 03 (ObsID 6155) in Very Faint telemetry format. In both observations the target was placed  $0\farcm15$  from the optical axis on ACIS-S3 chip, other ACIS chips were not activated. In ObsID 6154 a 101 pixels (49\farcs7) subarray was used (frame time 0.4 s), while the whole chip was read out in ObsID 6155 (frame time 3 s). We processed the data using the \emph{Chandra} Interactive Analysis of Observations (CIAO\footnote{See \url{http://cxc.harvard.edu/ciao/index.html}.}) software (ver.\ 4.6) and {\sl CXO} Calibration Data Base (CALDB ver.\ 4.5.9).

\section{Results}

\subsection{UV and optical images}

\begin{figure}
\centering
\includegraphics[scale=0.38]{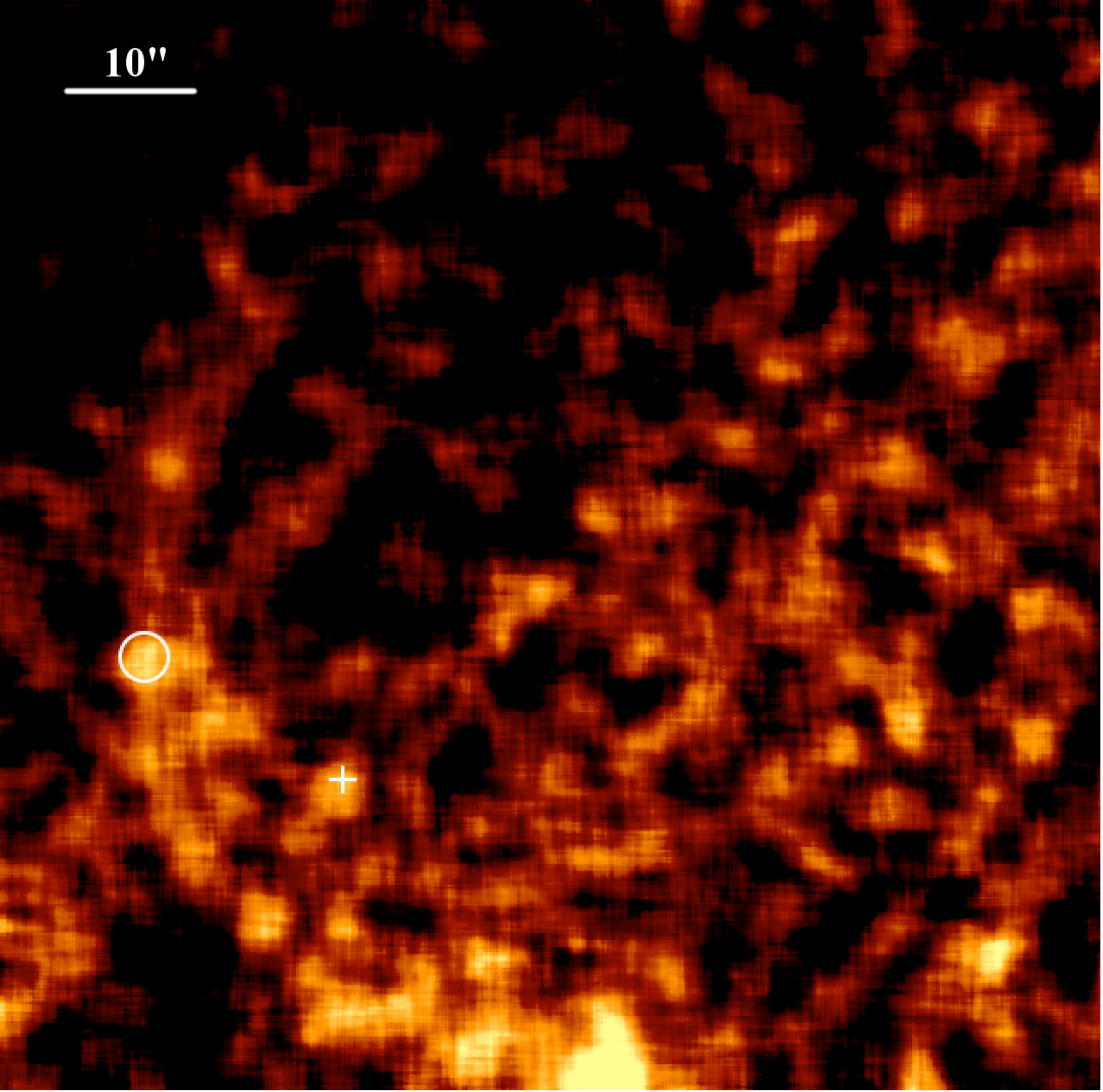}
\caption{ACS/WFC F658N image with median smoothing (within $r=0\farcs75$) applied. The FUV ``'blob'' (see Figure~\ref{panel}, mid-left panel) is within the $r=1\farcs6$ white circle. The pulsar position is shown by the cross.}
\label{Halpha}
\end{figure}

The FUV F140LP and F125LP images in Figure~\ref{panel} clearly reveal a bent extended structure ahead of the pulsar.  Its apex distance,  $\approx 10''$ ($2.3\times 10^{16}$\,cm), and shape are very similar to those of the H$\alpha$ bow shock. To compare the FUV and H$\alpha$ structures, we plotted H$\alpha$ surface brightness contours, within which the brighter part of the H$\alpha$ bow shock is confined in the middle-right panel of Figure~\ref{panel} (the selected contour represents the level for which the H$\alpha$ bow shock is not broken down into individual parts). Since the H$\alpha$ bow shock is only marginally detected in our  2013 ACS/WFC F658N image (Figure~\ref{Halpha}), we used an archival SOAR image obtained on 2012 March 21 (see top left panel of Figure~\ref{panel}) to produce the H$\alpha$ contours. Because the FUV structure very closely follows the H$\alpha$ structure, we conclude that it is an FUV counterpart of the H$\alpha$ bow shock. The apparent thickness of the FUV bow shock at its apex is about $2''$, somewhat smaller than that of the H$\alpha$ bow shock (but it depends on image depth). The FUV shock is also seen in the (dispersed) SBC PR130L image of 2006 (see bottom right panel of Figure~\ref{panel}), but no shock is detected in the HRC PR200L image, nor in the short HRC observations taken with the F300W and F555W filters in the same 2006 campaign.

In addition to the FUV bow shock, the F140LP image from the 2006 May 1 observation reveals a ``blob'' of $\approx$$3''$ size, overlapping with the bow shock limb (mid-left panel of Figure \ref{panel}). The  blob, centered at R.A.$= 4^{\rm h}37^{\rm m}16\fs94$, Decl.$= -47^\circ15'02\farcs0$, is clearly resolved but appears to be amorphous, with no clear structure. It is out of the SBC fields of view in the FUV images of 2013, but it is marginally detected in the 2013 WFC H$\alpha$ image  (Figure~\ref{Halpha}). The ratios of the surface brightness of the blob to that of the bow shock are $\approx$2.5 and $\approx$1.5 in the F140LP and H$\alpha$ images, respectively. Interestingly, while the bow shock was not detected in the 2006 HRC/PR200L data, the brighter UV blob was. This proves that the blob is not an  artifact of the SBC detector. Furthermore, the blob is not seen in the previous high-resolution H$\alpha$ images (e.g., in the image by BR14), which suggest its variability. Some of the archival optical/UV images (WFPC2 F555W and F814W, ACS/HRC F300W and F555W; see Table 1 for observation dates) show a faint point source near the eastern boundary of the F140LP blob but no trace of extended emission. It remains unclear whether the point source is physically connected with the blob or it is just a projection effect.

\subsection{Chandra images and spectrum}

The inspection of the combined {\sl Chandra} ACIS-S image (shown in Figure~\ref{panel}, top-right panel) reveals a faint diffuse X-ray emission extending up to about $5''$ south-east of J0437 (i.e., in the direction of pulsar's proper motion). For a conservative estimate of the detection significance, we extracted the source and background counts from the smaller and larger sectors of the $1\farcs87 < r < 4\farcs68$ annulus around the pulsar position, shown by green lines in the top right panel of Figure~\ref{panel}. We found $44\pm 7$ background-subtracted counts (110 and 230 in the source and background areas, respectively) in the area of 13.1 arcsec$^2$ of the smaller sector, i.e., the extended emission (presumably an X-ray PWN) is detected with a $>6\sigma$ significance. \citet{2002ApJ...569..894Z} could not detect such a small, faint nebula in their 25 ks ACIS-S observation of 2000 May 29 (ObsID 741) because of strong PSF degradation at the $3\farcm9$ off-axis target position (see Section 2.2), while the 20 ks HRC-I observation (2000 February 13; ObsID 742), also analyzed by these authors, was not deep enough for PWN detection.

We fit the spectrum of the detected extended emission with the absorbed power-law (PL) model with fixed $N_{\rm H}=7\times 10^{19}$ cm$^{-2}$ (this value was determined by \citealt{2002ApJ...569..894Z} from fits of the X-ray pulsar spectrum). The fit yields the photon index $\Gamma=1.8\pm0.4$ and absorbed flux $F_{0.5-8\,{\rm keV}}=(1.0\pm0.2)\times10^{-14}$\,erg\,s$^{-1}$\,cm$^{-2}$, which corresponds to luminosity $L_{\rm 0.5-8\,keV}\approx3\times10^{28}$\,erg\,s$^{-1}$ (at the distance of J0437).

\subsection{Spectral analysis of the FUV data}

We use two independent measurements to constrain the FUV  spectrum for  the J0437 bow shock: (1)  the F140LP and F125LP imaging photometry  obtained in the 2013 program; and (2)  the PR130L slitless spectroscopy from the 2006 program. For each of the two measurements we test two spectral models. First model is a simple PL continuum. It would be directly applicable to synchrotron emission from relativistic electrons with a PL spectral energy distribution, and it could crudely characterize the slope of a more complicated spectrum. Second model is an  FUV spectrum (spectral lines plus continuum) emitted from ISM matter compressed and heated in the forward shock region. Limited by the data quality,  we have to assume  that  the spectrum does not vary  throughout the observed bow shock region.

\subsubsection{F125LP and F140LP photometric data}

We selected the same source and background extraction regions for the F125LP and F140LP images (shown in Figure~\ref{panel}, middle-right and bottom-left panels, respectively). One of the two background regions was selected outside the bow shock to avoid the bow shock emission inside the bow. However, the background from this region (hereafter, 1-sided background) may be lower than the background at the bow shock location because of the nonuniform thermal glow contribution. Therefore, we also selected another background region inside the bow shock and used the ``inside + outside'' region for an alternative background estimate (hereafter, 2-sided background), which may somewhat exceed the true background because it includes a contribution from the shock emission.

\begin{deluxetable}{cccccc}
\tablecaption{Photometric data results \label{tbl-phot}}
\tablehead{
\colhead{Filter} & \colhead{$A_{s}$} & \colhead{$A_{b}$} & \colhead{$C_{s}$} & \colhead{$C_{b}$} \\
\colhead{} & \colhead{arcsec$^2$} & \colhead{arcsec$^2$} & \colhead{cnts s$^{-1}$}& \colhead{cnts s$^{-1}$}}
\startdata
F125LP\tablenotemark{a} & 32.2 & 29.4 & $1.33\pm0.04$ & $2.15\pm0.02$ \\
F125LP\tablenotemark{b} & 32.2 & 61.8 & $0.92\pm0.03$ & $5.32\pm0.03$ \\
F140LP\tablenotemark{a} &  32.2 & 29.4 & $0.82\pm0.03$ & $1.06\pm0.02$ \\
F140LP\tablenotemark{b} &  32.2 & 61.8 & $0.56\pm0.02$ & $2.75\pm0.03$
\enddata
\tablenotetext{a}{For 1-sided background (from the``outer'' region only).}
\tablenotetext{b}{For 2-sided background (from both the ``inner'' and ``outer'' regions);
$A_{\rm b}$ is the combined area of the two background regions.}
\end{deluxetable}

\begin{figure}
\includegraphics[scale=0.45]{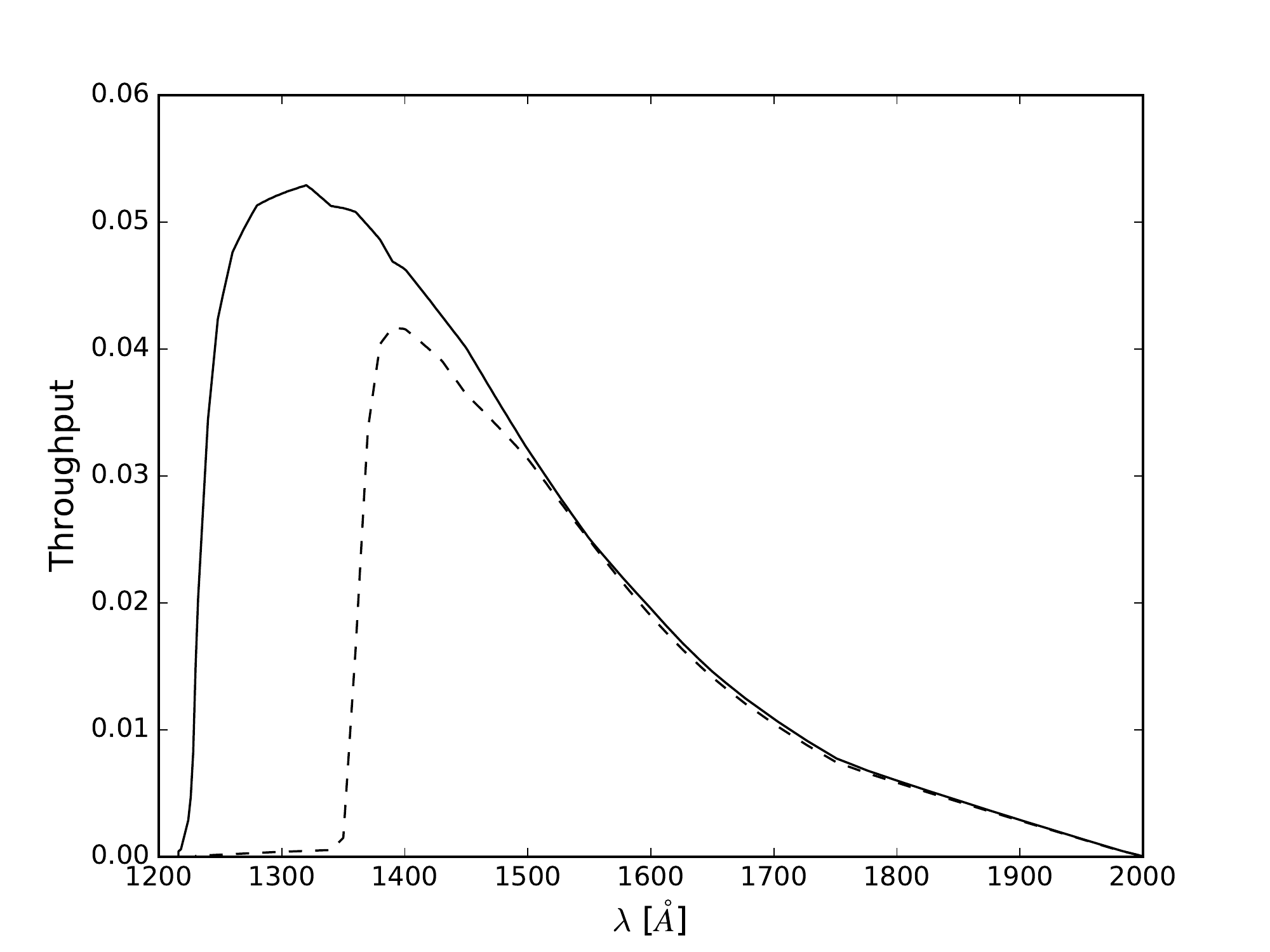}
\caption{SBC F125LP (solid line) and F140LP (dashed line) throughputs.}
\label{throughputs}
\end{figure}

The net source count rate for each background choice is calculated as $C_{s}=C_{t} - C_{b} (A_{s}/A_{b})$, where $C_{t}$ is the total count rate in the source region of area $A_{s}$, and $C_{b}$ is the background count rate in the region of area $A_{b}$. Correspondingly, the source count rate uncertainty is $\delta C_{s}=\left[ C_{t} + C_{b} (A_{s}/A_{b})^2 \right]^{1/2}t_{\rm exp}^{-1/2}$, where $t_{\rm exp}$ is the exposure time. The F125LP and F140LP count rates and their uncertainties for two background choices are given in Table~\ref{tbl-phot}.  The source count rates estimated with the 1-sided background exceed those estimated with the 2-sided background by a factor of about 1.45, but the ratio of the source count rates, $C_{\rm F125}/C_{\rm F140} = 1.62\pm 0.08$ and $1.64\pm 0.08$, is virtually the same for the two background choices. Since the throughputs of the two filters coincide at $\lambda\gtrsim 1400$\,\AA\ (see Figure \ref{throughputs}), the fact that $C_{\rm F125}/C_{\rm F140}>1$ means that both the 1250--1400\,\AA\ and 1400--2000\,\AA\ wavelength bands are contributing to the FUV bow shock radiation, i.e., the FUV emission is not concentrated in one spectral line. Below we will use the measured count rates to constrain the bow-shock FUV spectrum, with the aid of the PyRAF package {\tt PySYNPHOT}.

\begin{figure}
\includegraphics[scale=0.45, trim=20 20 0 0]{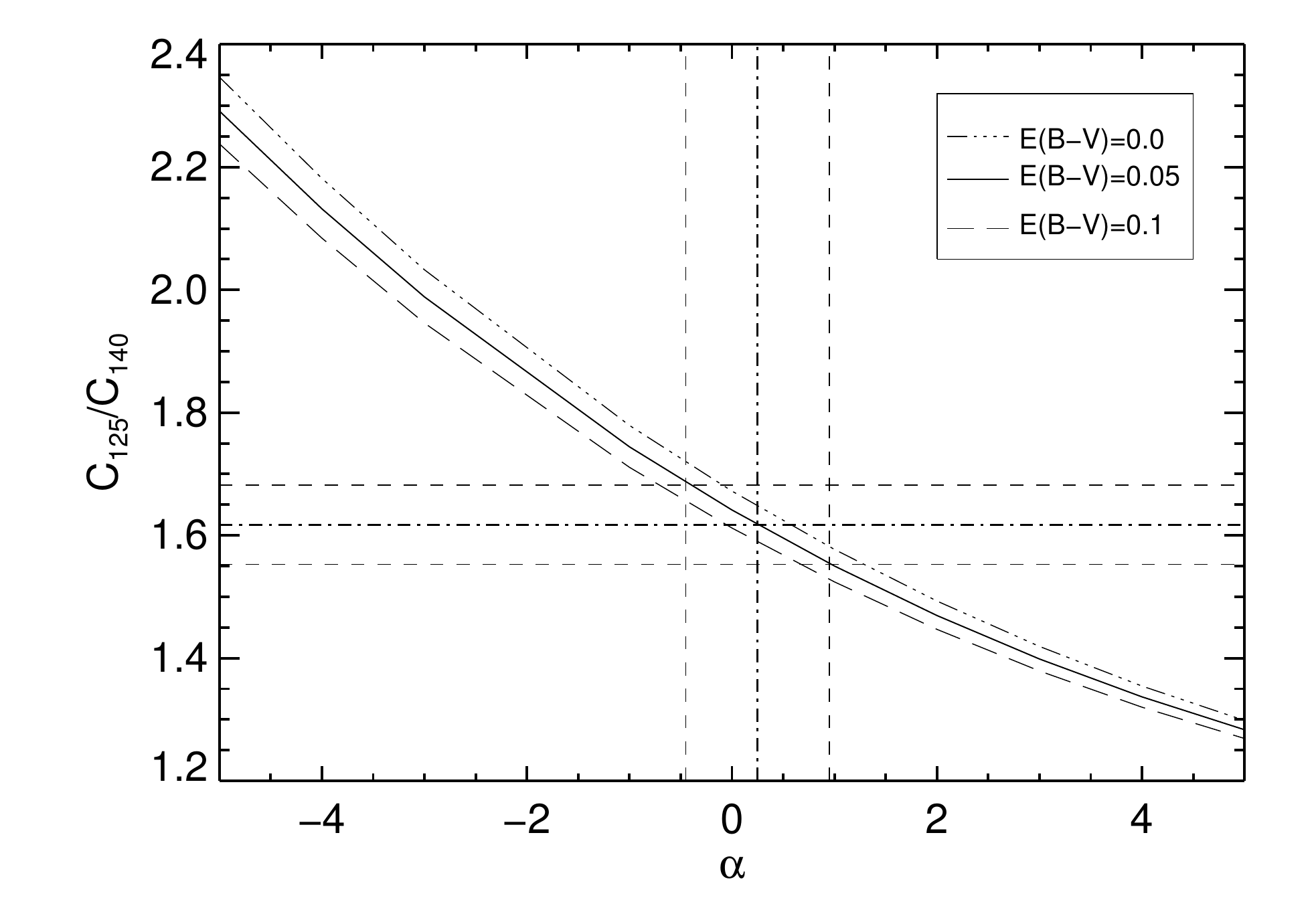}
\caption{
The ratio of the net count rates in the two filters (F125LP and F140LP) as a function of PL slope $\alpha$ ($F_\lambda\propto\lambda^{\alpha}$) for three different values of reddening, $E(B-V)= 0.01$, 0.05 and 0.1. The horizontal lines mark the measured value of the ratio of the net count rates and its uncertainties (for the 2-sided background; see Section 3.3.1 for details). The vertical lines show the range of allowed slopes for $0.01<E(B-V)<0.1$.
}
\label{slope}
\end{figure}

For the absorbed {\em PL model}, we simulate the count rates $C_{\rm F125}$ and $C_{\rm F140}$ as functions of slope $\alpha$ of the PL spectrum, $f_\lambda = f_{\lambda_0} (\lambda/\lambda_0)^{\alpha}$,  for three reddening values, $E(B-V)=0.01$, 0.05, and 0.1, in a plausible reddening range. Then we compare the simulated count rate ratio (which does not depend on  PL normalization) with the measured one, which allows us to determine the best-fit slope and its uncertainties for a given reddening (see Figure~\ref{slope}). We found that the best-fit $f_\lambda$ slope is nearly flat, $\alpha\approx 0$, corresponding to flux densities, $f_{\lambda_0}$, of $(1.58\pm0.05)\times10^{-17}$ and $(1.52\pm0.05)\times10^{-17}$\,erg\,s$^{-1}$\,cm$^{-2}$\,\AA$^{-1}$ (in the F125LP and F140LP filters, respectively), but the uncertainties of the spectral slope are large. For instance, $\alpha = 0.1\pm0.9$ and $\alpha=0.2\pm 0.7$ for the 2-sided and 1-sided background, respectively, at the most plausible $E(B-V)=0.05$ \citep{2012ApJ...746....6D}. Based on this model, we calculate the absorbed model fluxes in the F125LP and F140LP wavelength ranges (which only weakly depend on the assumed reddening): $F(1250$--$2000\,{\rm \AA})= (1.18\pm 0.04) \times10^{-14}$\,erg\,s$^{-1}$\,cm$^{-2}$ and $F(1350$--$2000\,{\rm \AA})= (1.03\pm 0.04) \times10^{-14}$\,erg\,s$^{-1}$\,cm$^{-2}$, for the 2-sided background. They correspond to luminosities $L\approx 5.0\times 10^{28}$ and $4.3\times 10^{28}$ erg s$^{-1}$, respectively. We should note that the quoted flux uncertainties are statistical ones; systematic uncertainties are substantially larger. In particular, all the fluxes and luminosities would be a factor of 1.45 larger if we use the 1-sided background. We also stress that these fluxes and luminosities are for the image area of 32.2 arcsec$^2$, which is likely a small fraction of the entire FUV bow shock area, so the full fluxes/luminosities can be higher.

The estimated FUV fluxes and luminosities are larger than the H$\alpha$ ones from the same region of the bow shock. A crude estimate from our shallow {\sl HST} H$\alpha$ image (Figure \ref{Halpha}) gives the photon flux  ${\cal F}_{{\rm H}\alpha}\sim 0.4\times 10^{-3}$\, cm$^{-2}$\,s$^{-1}$ in the FUV extraction region (the value strongly depends on background region choice). Alternatively, the H$\alpha$ flux can be estimated by scaling the BR14 flux, $6.7\times 10^{-3}$\,cm$^{-2}$\,s$^{-1}$, measured in the larger ``apex zone'', using the ratio ($\simeq 13)$ of source counts in the apex zone to that in the FUV extraction region projected onto the SOAR image. This gives ${\cal F}_{{\rm H}\alpha}\approx 0.5\times 10^{-3}$\, cm$^{-2}$\,s$^{-1}$, slightly higher than (but within the uncertainties of) the flux estimated from the {\sl HST} H$\alpha$ image. The latter photon flux value, which we consider more reliable, corresponds to the energy flux $F_{{\rm H}\alpha} = 1.5\times 10^{-15}$\,erg\,cm$^{-2}$\,s$^{-1}$, a factor of 
8 (11) lower than the absorbed (unabsorbed) flux in the 1250--2000\,\AA\ range.

We can also use the PL model to estimate the flux of the blob detected in the F140LP image. Assuming a flat $f_\lambda$ spectrum, we obtain $F_{\rm blob}(1350-2000\,{\rm \AA})=(0.66\pm 0.03)\times 10^{-14}$ erg cm$^{-2}$ s$^{-1}$ (observed flux), which corresponds to the luminosity $L_{\rm blob}\sim 2\times 10^{28}$ erg s$^{-1}$. It is larger than the blob's flux/luminosity in the F658N H$\alpha$ filter by a factor of about 15 (this factor depends on choice of the background region).

The other model we use to compare with the observed FUV bow shock properties is the {\em emission spectrum from shocked ISM matter}, compressed and heated while passing through the shock. Numerical models of collisionless shocks and their emission have been developed by many authors (see, e.g., the review by \citealt{2013SSRv..178..599B} for references). However, as we are not aware of {\em bow shock} FUV emission models in the regime appropriate for J0437, we have to use models for a {\em plane shock}. For model calculations, we use the {\tt SHELLS} code described in Section 5 of the \citet{2013SSRv..178..599B} review. This code allows one to calculate radiation spectra (continuum and lines) of one-dimensional shocked flows in the far-infrared$-$FUV range, simultaneously with the upstream and downstream flow structure for a collisionless shock moving with a $v_s=80$--400 km\,s$^{-1}$ velocity through the ISM of an $n_0= 0.1$--25\,cm$^{-3}$ upstream number density ($n_0=\rho/\mu_i m_H$, where $\rho$ is the mass density, and $\mu_i$ is the mean atomic mass of ions). The model accounts for radiative cooling and heating, non-equilibrium photoionization and recombination, various emission mechanism (bound-bound, free-bound, free-free, and two-photon transitions), radiation transfer, and magnetic fields.

\begin{figure}
\centering
\includegraphics[scale=0.6]{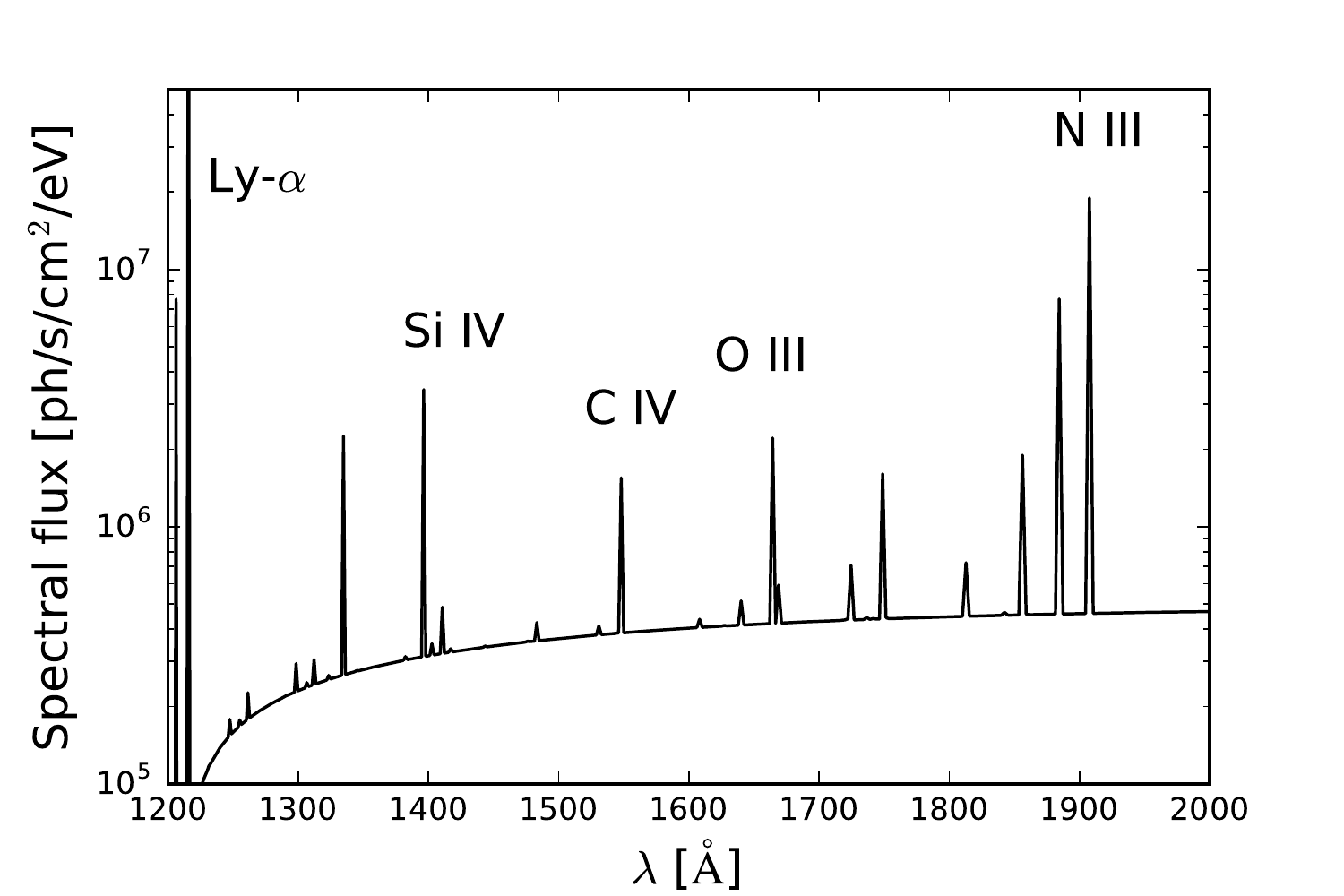}
\caption{
Model FUV spectrum emitted from 1\,cm$^2$ of the plane shock with $v_s=110$\,km\,s$^{-1}$, $n_0=0.2$\, cm$^{-3}$. Some of the lines are labeled.}
\label{shock-model}
\end{figure}

\begin{figure}
\hspace{-0.5cm}
\includegraphics[scale=0.4]{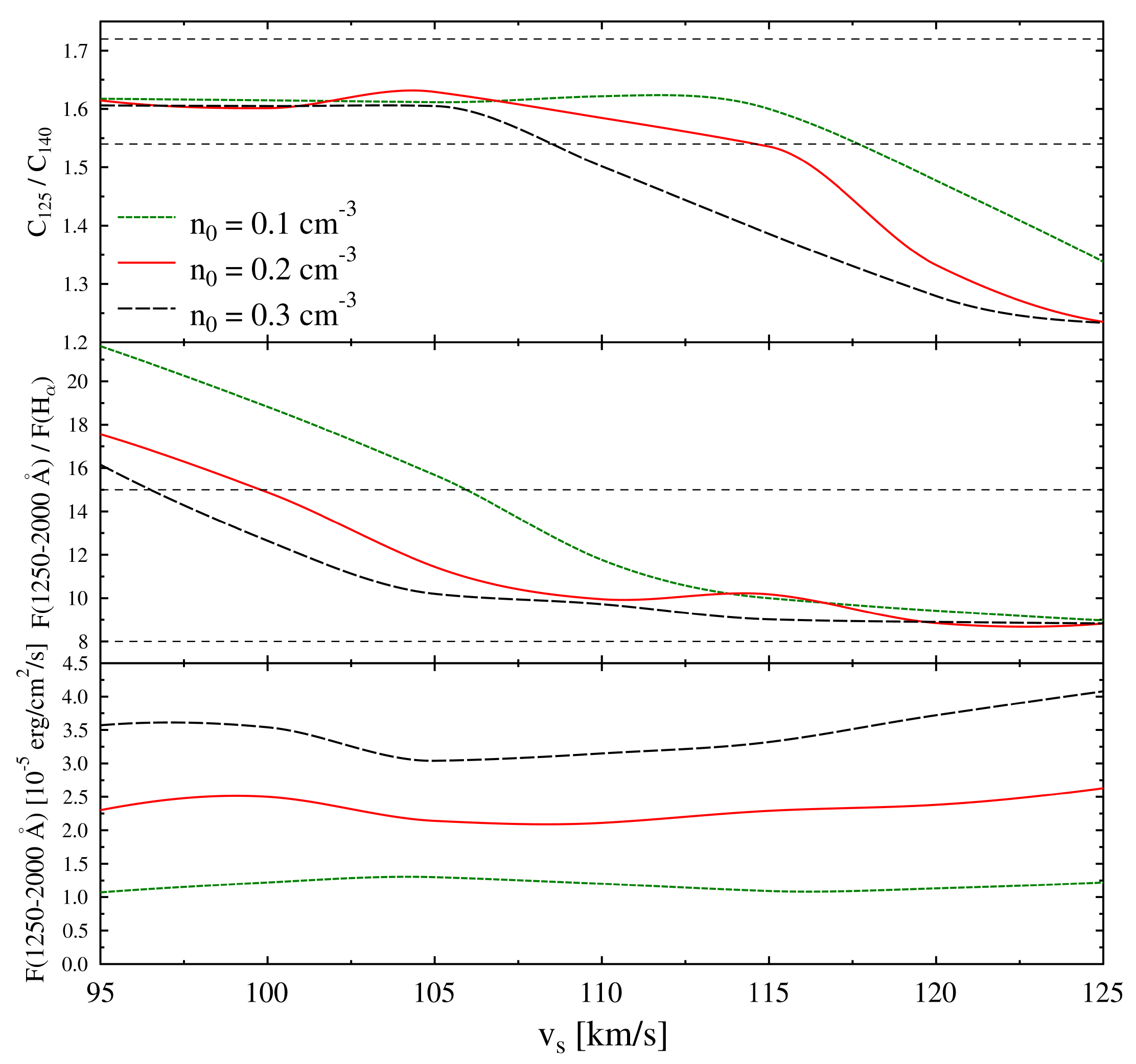}
\caption{Count rate ratio $C_{\rm F125}/C_{\rm F140}$ for $E(B-V)=0.05$ (top), $F(1250$-$2000\,{\rm \AA})/F_{{\rm H}\alpha}$ ratio (middle; the fluxes are corrected for extinction) and the model flux emitted from unit area of plane shock, $F_{\rm mod}(1250$-$2000\,{\rm \AA})$ in units of $10^{-5}$\,erg\,cm$^{-2}$\,s$^{-1}$, as functions of shock speed $v_s$ for preshock densities 0.1, 0.2, and 0.3 cm$^{-3}$, computed with the {\tt SHELLS} code. The dotted horizontal lines in the top and middle panels show the observational bounds on the corresponding quantities.
}
\label{shells_fluxes}
\end{figure}

For the comparison with our observations, we computed a set of {\tt SHELLS} models for different combinations of the most important parameters $v_s$ and $n_0$, assuming a solar chemical composition  ($\mu_i=1.242$) and an upstream (ISM) magnetic field of $3\,\mu{\rm G}$ perpendicular to the shock front. An example of the FUV part of the spectrum, for $v_s=110$\,km\,s$^{-1}$, $n_0=0.2$\,cm$^{-3}$, is shown in Figure \ref{shock-model}. For each of these combinations, we calculated the expected count rate ratios $C_{\rm F125}/C_{\rm F140}$ with the aid of {\tt PySYNPHOT} package, for plausible reddening $E(B-V)$, and compared them to the observed ratio, similarly to the absorbed PL model. The upper panel of Figure \ref{shells_fluxes} shows that at $E(B-V)=0.05$ the model count rate ratio becomes smaller than the observed range, $C_{\rm F125}/C_{\rm F140}=1.54$--1.72, for $v_s \gtrsim 109$--118\,km\,s$^{-1}$, depending on preshock density in a plausible range $n_0=0.1$--0.3\, cm$^{-3}$ (see Section 4).

\begin{figure}
\hspace{-0.2cm}
\includegraphics[scale=0.58]{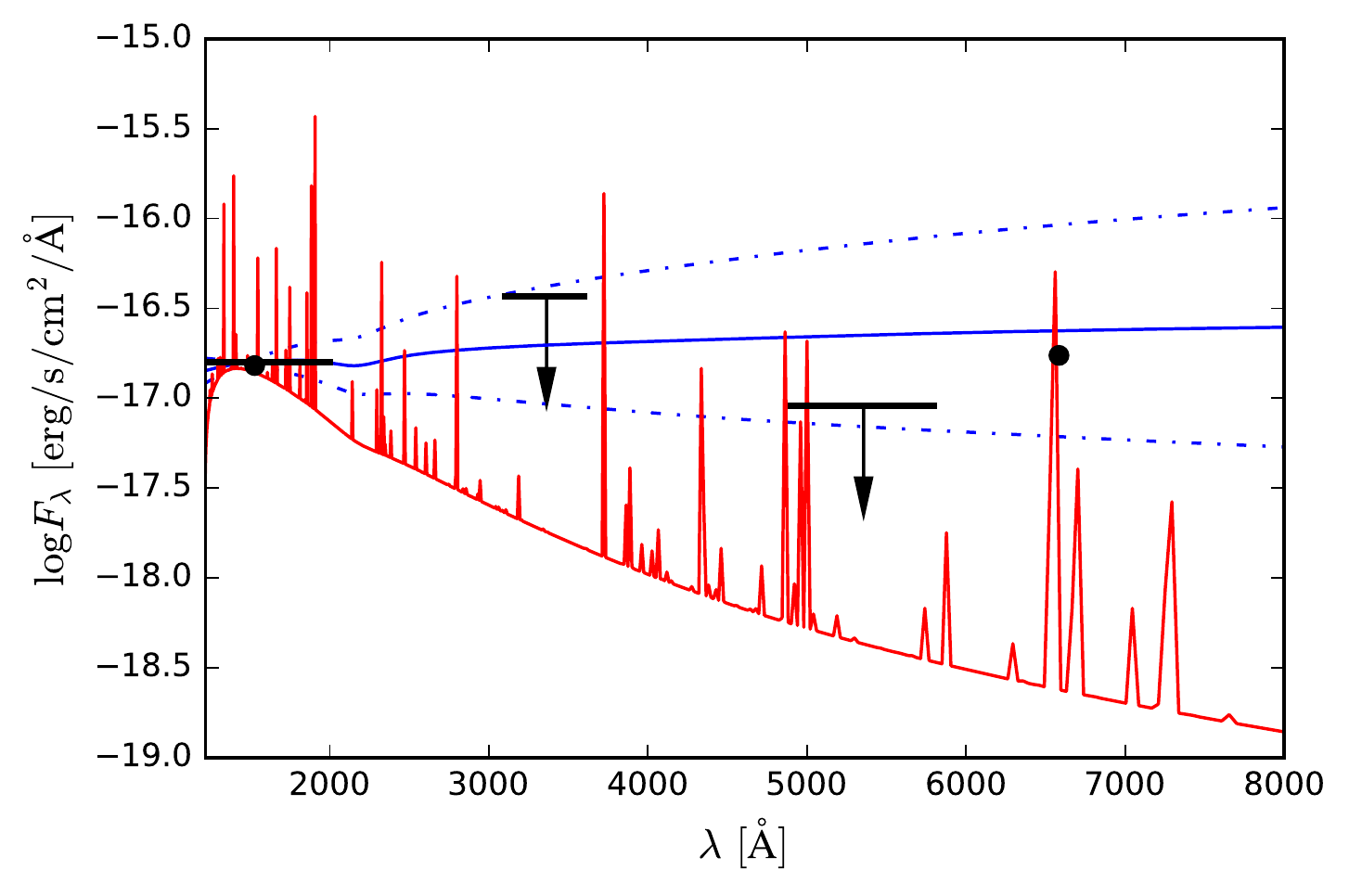}
\caption{
Measured flux density values and upper limits in the bow shock region (see Figure~\ref{panel}) are shown together with spectral models. The black dots show the F125LP and F658N spectral fluxes (the narrower F140LP band is contained within the F125LP one and hence it is not shown for clarity). The arrows show the $3\sigma$ upper limits for the HRC F330W and F555W observations calculated assuming a flat $F_\lambda$ spectrum.. The horizontal black lines represent the filter widths at the half-maximum. Absorbed flux densities for the PL ($\alpha=0.1\pm0.9$) and shocked plasma model (for $v_s=110$\,km\,s$^{-1}$ and $n_0=0.2$\,cm$^{-3}$) are plotted as blue and red lines, respectively. The dot-dashed blue lines represent the $1\sigma$ uncertainty in the PL slope.
}
\label{limits}
\end{figure}

Another model parameter to compare with observations is the ratio of the FUV and H$\alpha$ fluxes. The middle panel of Figure \ref{shells_fluxes} shows that the ratio of the model energy fluxes $F(1250$--$2000\,{\rm \AA})/F_{{\rm H}\alpha}$ increases with decreasing $v_s$ and becomes larger than the observed flux ratio, $\approx 8$--15 (corrected for extinction $E(B-V)=0.05$), at $v_s\lesssim 97$--106\,km\,s$^{-1}$, depending on $n_0$. Thus, comparing the flux ratios in 3 filters, we constrain the shock speed: $97\lesssim v_s\lesssim 118$\,km\,s$^{-1}$, at $0.1\leq n_0\leq 0.3$\,cm$^{-3}$ ($100\lesssim v_s\lesssim 115$\,km\,s$^{-1}$ at $n_0=0.2$ cm$^{-3}$). We note that these constraints are derived for the effective {\em plane shock} velocity, which can be lower than the bow-shock (pulsar) velocity because the velocity component perpendicular to the bow shock is lower than the pulsar velocity except for the bow-shock apex.

Although the bow shock was not detected in the F300W and F555W images, it is interesting to compare upper limits at these wavelengths  with the spectral models. Due to the apparent non-uniformities\footnote{These non-uniformities are particularly strong in the F330W image which suffers from cosmic ray contamination due to the lack of dithering or CR splits.} in the background, clearly exceeding the statistical fluctuations, we chose to measure the upper limits by sampling\footnote{The areas of the background regions were equal to that of the bow shock region used for FUV flux measurements.} the background counts from five different regions in the vicinity of the bow shock. The standard deviation of these measurements was used as a $1\sigma$ upper limit. The corresponding $3\sigma$ upper limits of the flux densities are $4\times10^{-17}$ and  $9\times10^{-18}$\,erg\,s$^{-1}$\,cm$^{-2}$\,\AA$^{-1}$ in the F330W and F555W images, respectively. Figure~\ref{limits} shows the best-fit PL and shocked plasma models together with the $3\sigma$ limits and the FUV and H$\alpha$ flux density measurements. We see that the optical upper limits are too high to distinguish between the PL and shocked plasma models, but the F555W limit favors smaller values of the PL slope, $\alpha\lesssim -0.6$ in the broad range of $\alpha$ obtained from the FUV data for the PL model (Section 3.3.1).

\subsubsection{PR130L prism data}

A small part of the FUV bow shock was also detected with the SBC/PR130L prism (see Figure 1, bottom-right panel). Even though the SBC prism is designed for spectroscopy of point sources, we attempted to analyze the prism data to check if any additional information about the bow shock spectrum can be obtained. 

\begin{figure}
\centering
\includegraphics[scale=0.55]{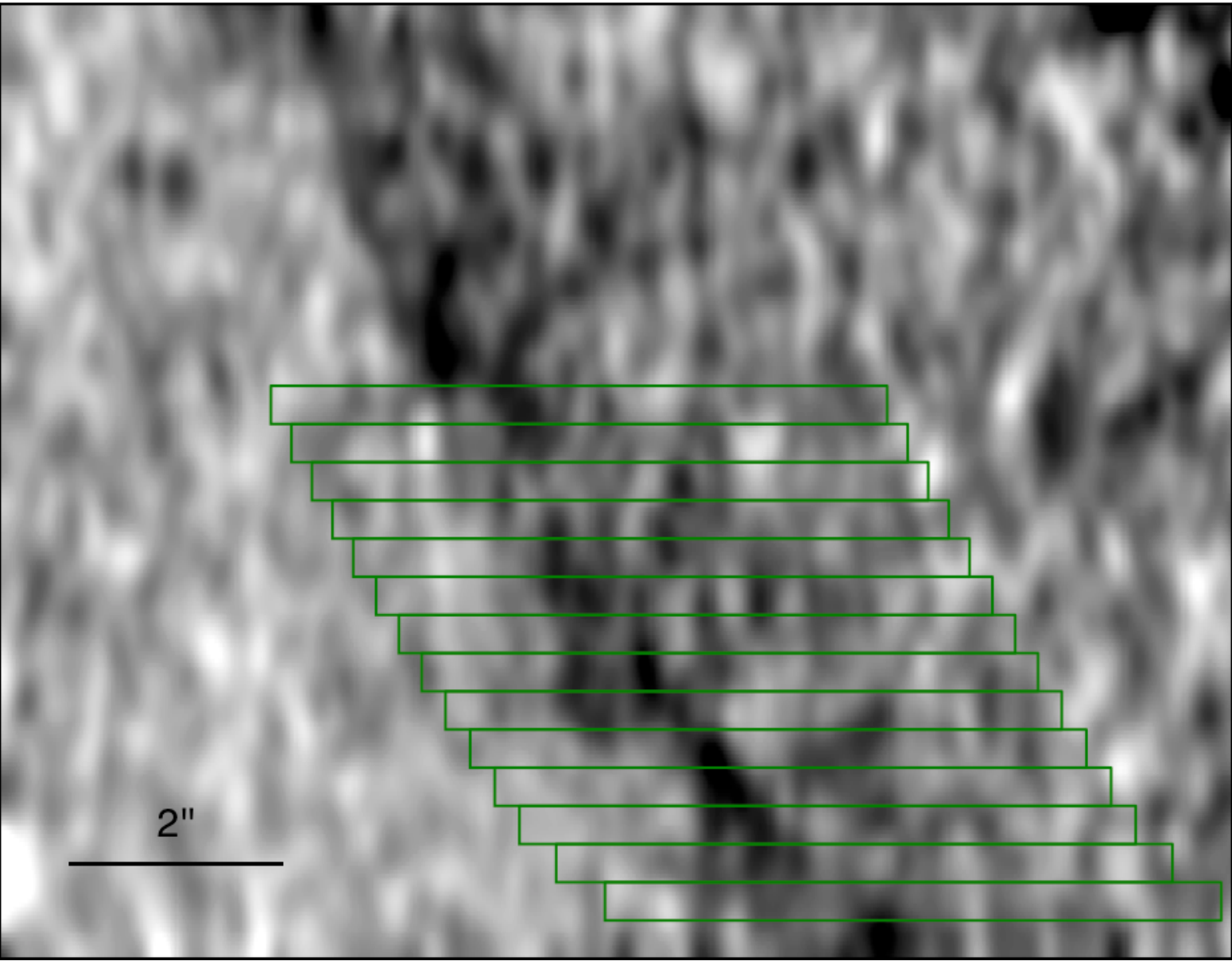}
\caption{The dispersed PR130L image of the part of the bow shock with stripe regions used for spectral extraction. The prism dispersion is in the horizontal direction.}
\label{prism}
\end{figure}

\begin{figure*}
\centering
\includegraphics[scale=0.441]{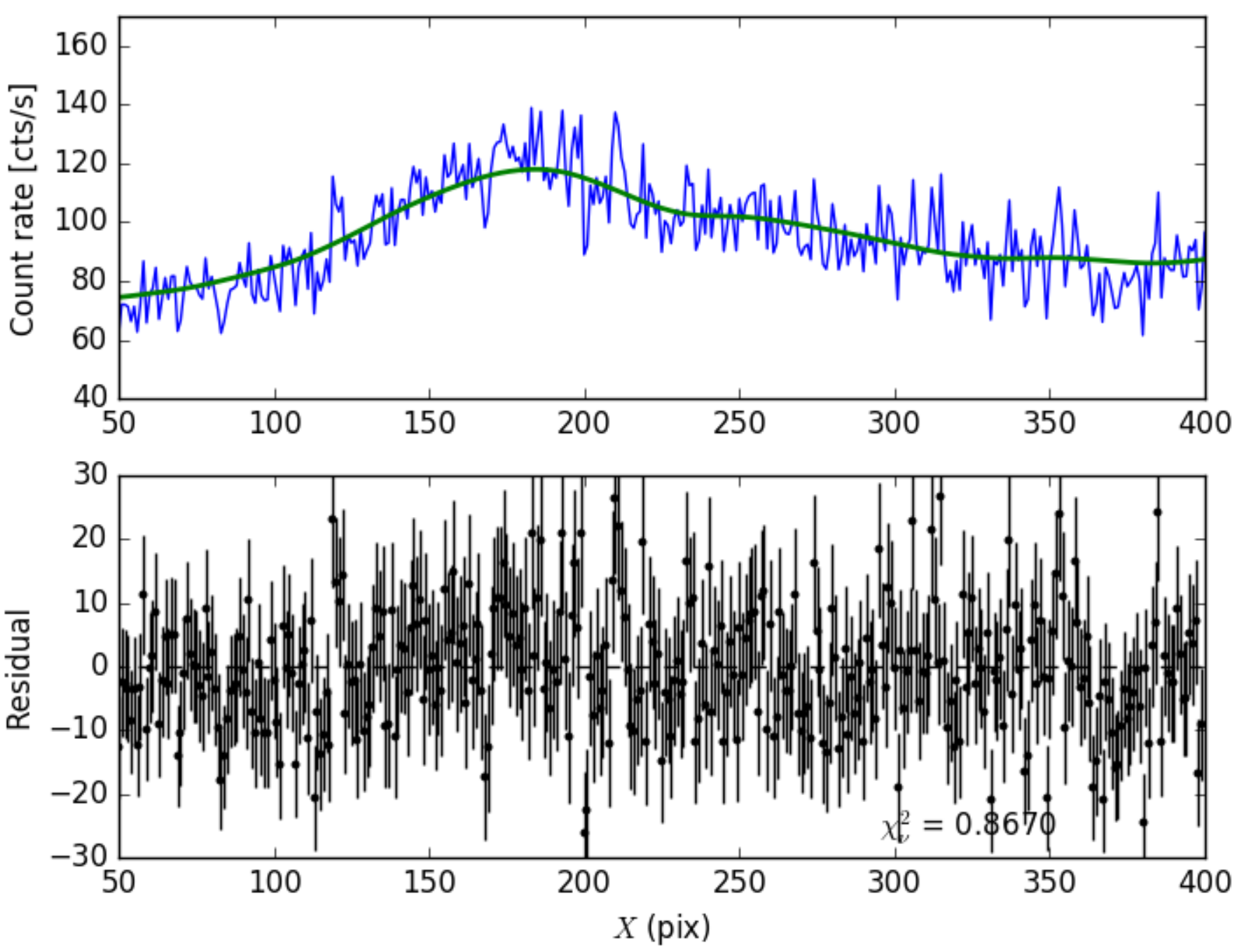}
\includegraphics[scale=0.441]{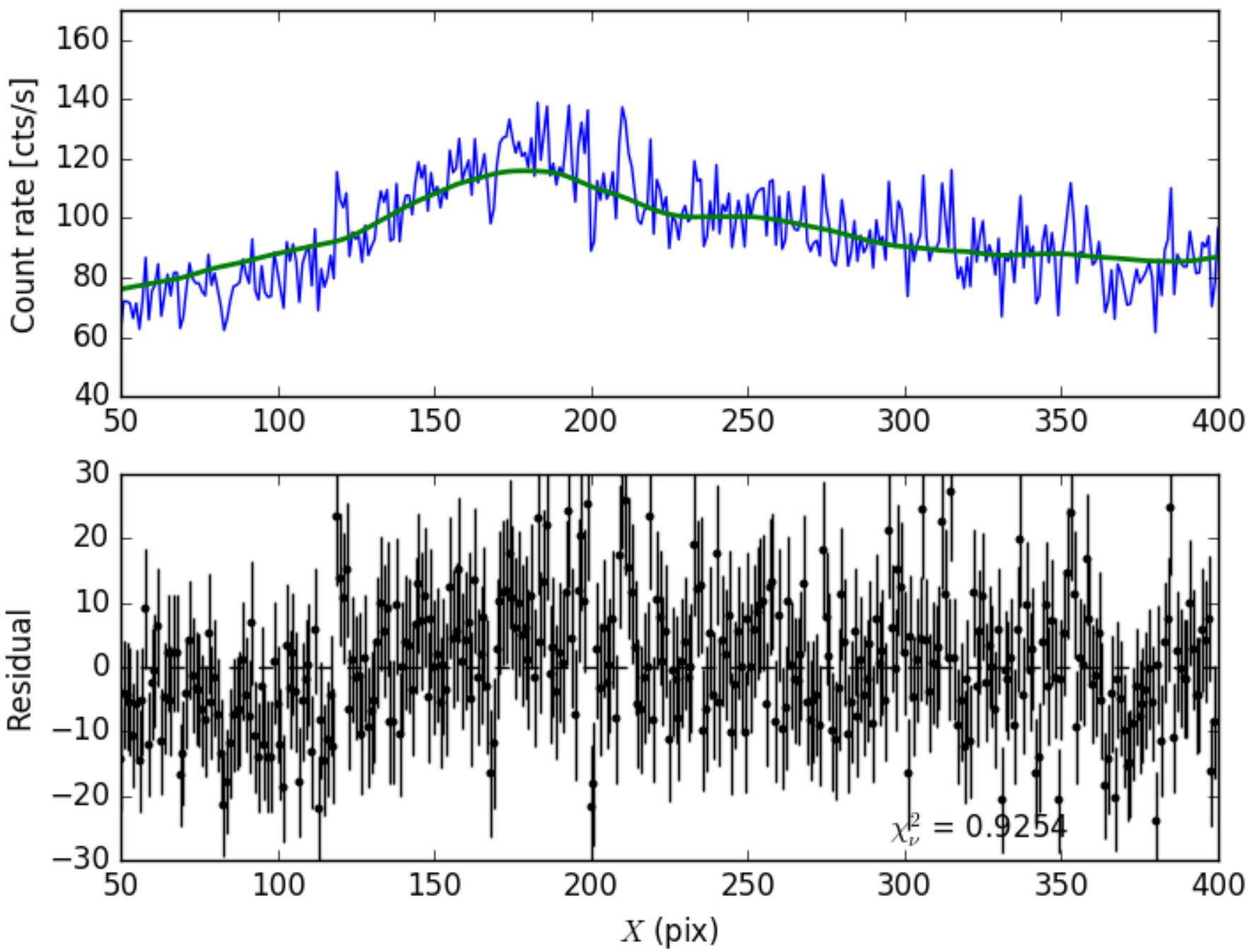}
\caption{
The left and right panels show the best-fit PL model ($\alpha=0$)  and collisionless shock emission model ($v_s=110$\,km\,s$^{-1}$, $n_0=0.2$\,cm$^{-3}$), respectively (see Section 3.3.2 for details).}
\label{model}
\end{figure*}

Since the bow shock is an extended feature, we had to stack the data
in the cross-dispersion (vertical) direction before we could fit the spectrum. We shifted the image data (using 10-pixel-wide  stripes; see Figure~\ref{prism}) following the curvature of the bow shock, and summed the counts along the vertical direction, to make a one-dimensional dispersed spectrum. We halted the extraction where the dispersed image nears the detector edge. Because the bow shock is resolved in the direct image along the dispersion (horizontal) direction, the stacked one-dimensional spectra are relatively broad (significantly broader than the spectrum of a point source would be). The observed dispersed spectrum of the bow shock is thus a convolution of the spatial profile of the bow along the dispersion direction with the known throughput\footnote{See \url{http://www.stsci.edu/hst/acs/analysis/STECF/isr/isr0602.pdf}} and  dispersion of the prism. To fit the observed spectrum with a model, we follow a forward-fitting method, i.e., calculate the dispersed PR130L spectrum (in detector coordinates) for a point source, convolve it with the bow shock profile measured from the direct image, and compare with the observed prism profile.

While stacking multiple spectra (Figure~\ref{prism}) increases the signal-to-noise ratio ($S/N$), it also complicates the background subtraction. Unlike a point source extraction, where the background can be estimated in the immediate vicinity of the source, the background for each spectral extraction region (stripes in Figure~\ref{prism}) is slightly different. Moreover, the thermal glow also changes across different parts of the detector. In addition, the profile from the direct image (which is used for the model spectrum convolution) comes from different detector coordinates than the profile from the dispersed prism image, where the background and thermal glow are different. Finally, any changes in the bow shock profile or spectrum along the bow shock are not accounted for in our approach. This complicates the uncertainty estimation of our final results, and these systematic errors significantly exceed formal statistical errors obtained during the model fitting. We explored various background subtraction procedures, including a constant background and a polynomial (with background levels rising across the dispersion direction). Since in the interior of the bow shock we cannot distinguish between the background due to the SBC thermal glow and the real emission from the bow shock itself, we decided to adopt a flat background determined from a region outside of the bow shock. Although the  emission is present throughout the entire sensitivity range of PR130L (1100$-$1800\,\AA), the dispersed spectrum overlaps with the thermal glow. Therefore, we restrict our fitting to the 100$-$300\, pixel range (Figure~\ref{model}), corresponding to the stripes shown on Figure~\ref{prism}.

We find that under these assumptions the data can be satisfactory described by PL models  with $-1\lesssim \alpha \lesssim 1$ (Figure~\ref{model}, left panel), which is consistent with the $\alpha$ values obtained from the observations with broad filters.

Again, as an alternative to the simple PL model, we also used a model of emission from a  collisionless plane shock \citep{2013SSRv..178..599B}. Through multiple trials, we established that the models with strong individual lines produce too narrow profiles that cannot fit the observed one. We concluded that models with weak lines, such as the model with $v_s=110$\,km\,s$^{-1}$, $n_0=0.2$\,cm$^{-3}$ (Figure \ref{shock-model}), are preferred by the fits (Figure \ref{model}, right panel). The results are not  sensitive to the extinction in the $E(B-V)=0$$-$0.05 range.

Thus, the poor statistics and  smearing due to the extended nature of the  bow shock do not allow us to discriminate between the PL and collisionless shock emission  models in the prism spectra, but at least models in which emission is dominated by a few individual lines can be excluded. Sensitive slit spectroscopy or integral field spectroscopy would be needed to reach further progress.

\section{Discussion}

Our FUV observations of J0437 have revealed a structure whose shape matches closely (within 0\farcs5 for the leading edge) the brighter part of the H$\alpha$ bow shock (see Figure~\ref{panel}). This strongly suggests that the FUV emission is due to the interaction of the pulsar wind with the ambient ISM and most likely comes from the shocked ISM matter. 

The shocked ISM matter is confined between the forward shock (FS) and the contact discontinuity (CD), which separates the shocked ISM from the shocked pulsar wind. For an isotropic pulsar wind, the distance $r_{\rm fs}$ from the pulsar to the FS apex is a factor of $\approx 1.3$--1.4 larger than the distance $r_{\rm cd}$ to the CD apex, which can be estimated as 
\begin{equation}
r_{\rm cd} = [\dot{E}/(4\pi \rho v^2c)]^{1/2},
\end{equation}
where $\rho = n_b m_{\rm H}$ is the ambient mass density, $n_b$ the barion number density, $v=v_\perp/\sin i$ the pulsar velocity\footnote{Rigorously speaking, the pulsar velocity relative to the ambient medium, $\vec{v}_{\rm psr} = \vec{v}+\vec{v}_{\rm orb}$, depends on the orbital phase. This should lead to periodic variations of the bow shock position with the period $P_{\rm bin}=5.74$\,d. The amplitude of these variations depends on the angle between the velocity $\vec{v}$ of the binary system and the orbital plane. Since this angle is unknown, and the pulsar's orbital velocity, $v_{\rm orb}\approx 19$ km s$^{-1}$, is small compared to the velocity of the binary, $v_{\rm orb}/v < 0.18$, we neglect this effect in our estimates.}, and $i$ is the angle between the pulsar velocity and the line of sight (e.g., \citealt{2005A&A...434..189B}). For the J0437 bow shock, $r_{\rm fs}\approx2.3\times 10^{16}$\,cm ($10''$ in the image) corresponds to $r_{\rm cd}\approx 1.7\times 10^{16}$\,cm ($7''$--$8''$ in the image), in agreement with the observed bow shock thickness near apex. Using the approximation by \citet{2005ApJ...629..979L}, $I_{45}\approx (M_{\rm PSR}/M_\odot)^{1.5}\approx 1.7$, we estimate the spin-down power as $\dot{E}\approx 5.0\times 10^{33}$\,erg\,s$^{-1}$, and obtain an  estimate for the ambient baryon density from Equation (1): $n_b \sim 0.25 \sin^2i$\,cm$^{-3}$, in agreement with the $n_{\rm H}\sim 0.2$\,cm$^{-3}$ estimated from the H$\alpha$ bow shock flux (BR14).

From the analysis of the H$\alpha$ bow shock emission, BR14 derived a high neutral H fraction, $\xi_{\rm HI}\approx 0.9$, in the preshock ambient medium, which implies a low degree of photoionization by UV and soft X-ray photons emitted from the pulsar, its PWN, and the shocked ISM. If the pulsar were not moving, it would ionize the ambient medium up to distances much larger than $r_{\rm fs}$ (within its Str\"{o}mgren sphere). However, as shown by \citet{1995ApJ...454..370B} (see also \citealt{2001A&A...380..221V} and \citealt{2015MNRAS.454.3886M}), the characteristic  size $r_0$ of the photoionized region ahead of a {\em moving} source of ionizing radiation can be estimated as
\begin{equation}
r_0 = (4\pi v)^{-1} \int_{I/h}^\infty \dot{N}_{\rm ph}(\nu) \sigma_{\rm ph}(\nu)\,d\nu\, ,
\end{equation}
where $v$ is the source (pulsar) velocity, $I=13.6$ eV is the hydrogen ionization potential, $\dot{N}_{\rm ph}(\nu)$ is the photon spectrum of ionizing radiation,
\begin{equation}
\sigma_{\rm ph}(\nu)=\frac{64\pi}{3 \sqrt{3}} \alpha_f a_{\rm B}^2 G(\nu) \left( \frac{I}{h\nu} \right)^3
\end{equation}
is the photoionization cross section, $\alpha_f$ is the fine-structure constant, $a_{\rm B}$ is the Bohr radius, and $G(\nu)$ is the Gaunt factor ($G(\nu)\approx 0.8$ at $h\nu \to I$). For J0437, the main source of ionizing radiation is thermal emission from the neutron star surface. We derive $\int_{I/h}^\infty \dot{N}_{\rm ph}(\nu) \sigma_{\rm ph}(\nu)\,d\nu = 3.2\times 10^{22}$ cm$^2$ s$^{-1}$ for the J0437 spectrum, which is well fitted with superposition of three blackbody components \citep{2012ApJ...746....6D}.
This gives $r_0 = 2.5\times 10^{14}\,\sin i$ cm, two orders of magnitude
smaller than $r_{\rm fs}$. Thus, the preshock degree of ionization is essentially the same as in the unperturbed ISM, thanks to the high pulsar speed. 

From our observations we estimated the FUV bow shock luminosity $L(1250-2000\,{\rm \AA})\approx 5\times 10^{28}$ erg s$^{-1}$, a factor of 10 higher than the H$\alpha$ luminosity from the same (small) region of the bow shock. The shape of the FUV spectrum remains poorly constrained because the imaging observations were done with only two very broad filters with overlapping passbands, while the prism observation was not  suited for spectroscopy of extended sources. Our analysis has shown, however, that the limited data available are consistent with model spectra of shocked ISM.  The ISM matter is compressed, heated, and partly ionized at the FS front. At a highly supersonic pulsar velocity the ions are heated up to $T_i\approx(3/16) (\mu_i m_p/k) v_{\rm psr}^2=2.3\times10^{5}\mu_i (v_{\rm psr}/100~{\rm km~s}^{-1})^2$~K, where $\mu_i m_p$ is the mean ion mass. Behind the FS front, the ions' energy is partly transferred to electrons and the flow speed decreases. Therefore, at some postshock distance the plasma temperature and density become high enough to emit a complex UV-optical spectrum comprised of both continuum and line emission.

Using the \texttt{SHELLS} code for plane-parallel shocks, we found that the observed count rate ratio in two FUV filters and the ratio of the FUV and H$\alpha$ fluxes are consistent with the model predictions at reasonable values of the preshock density and shock velocity (Section 3.3.1), and the model spectrum is consistent with the observed prism spectrum at the same parameters (Section 3.3.2). In addition to comparing the count rate and flux ratios (see Figure \ref{shells_fluxes}), a comparison of the observed shock fluxes with the model fluxes allows us to estimate the emitting area of the plane shock needed to provide the observed flux. Since the model flux is calculated per unit area of the plane shock, the emitting area can be estimated as $A=4\pi d^2 (F_{\rm obs}/F_{\rm mod}) = 2.9\times 10^{33} (F_{\rm obs,-14}/F_{\rm mod,-5})$\,cm$^2$, where $F_{\rm obs,-14}$ and $F_{\rm mod,-5}$ are the observed and model fluxes in units of $10^{-14}$ and $10^{-5}$\,erg\,cm$^{-2}$\,s$^{-1}$, respectively. The model flux in the 1250--2000\,\AA\ range is shown in the lower panel of Figure \ref{shells_fluxes} as a function of $v_s$ for three $n_0$ values. The flux weakly depends on $v_s$, and it is approximately proportional to the upstream density, $F_{\rm mod,-5}(1250-2000\,{\rm \AA})\approx 2(n_0/0.2\,{\rm cm}^{-3})$. The observed FUV flux, estimated with {\tt PySYNPHOT} for the calculated model spectra, weakly depends on the assumed model. Its characteristic value is $F_{\rm obs,-14}(1250-2000\,{\rm \AA})\approx 1.5$, and its uncertainty is mainly due to systematic errors. Using this estimate, we obtain $A\sim 2\times 10^{33}(n_0/0.2\,{\rm cm}^{-3})^{-1}$\, cm$^2$. This area corresponds to the radius $R=(A/\pi)^{1/2}\sim 2.5\times 10^{16} (n_0/0.2\,{\rm cm}^{-3})^{-1/2}$ cm (angular radius of $\sim 11''$ for $n_0=0.2$\,cm$^{-3}$) comparable with the FUV extraction region size perpendicular to the pulsar's proper motion direction. We should note, however, that the size of the downstream FUV emitting region in the plane shock model, $\sim$ a few $\times 10^{16}$\,cm for the plausible $v_s$ and $n_0$ values, considerably exceeds the thickness of the observed bow shock, $\sim5\times 10^{15}$\,cm, and even the stand-off distance of the bow shock apex, $\sim2\times 10^{16}$\,cm. This is not surprising because, contrary to the bow shock, the downstream region of the plane shock is not confined by the pulsar wind ram pressure, but it shows that applicability of plane shock models to bow shocks is limited. Thus, to prove that the observed FUV structure is indeed emission from a shocked ISM matter, the observational data should be compared to a realistic radiative bow shock model.

In some of our observations we detected an extended ``blob'' of $3''$ (460\,AU at $d=157$\,pc) size at the bow shock limb (Section 3.1). The lack of the blob in the wide-band  optical/NUV  images implies a spectrum that would be very unusual for a galaxy of any kind. The positional coincidence with the bow shock limb and the fact that the blob is also marginally detected in the ACS/WFC image from 2013 suggest that the blob is likely associated with the bow shock (e.g., caused by an ISM inhomogeneity or an instability in the FS region).

According to PWN models for supersonically moving pulsars (e.g., \citealt{2005A&A...434..189B}), the emission from the shocked ISM behind the FS should be accompanied by synchrotron emission from the shocked pulsar wind confined between the bullet-like termination shock (TS) surface and the CD surface. The synchrotron PWN is expected to have a cometary shape, with a ``head'' in front of the moving pulsar and a ``tail'' behind the pulsar. Such head-tail PWN morphologies have been observed in X-rays from a number of ordinary and recycled pulsars (see \citealt{2008AIPC..983..171K} for examples). Our analysis of archival X-ray observations of J0437 has shown an extended emission up to a distance of about $5''$  ahead of the pulsar (a factor of $\sim$1.5 smaller than the $r_{\rm cd}$ estimate)  but no tail behind. We presume that the detected emission  represents a brighter part of the PWN head\footnote{Such brighter emission can be seen in the simulated synchrotron map in Figure~4 of \citealt{2005A&A...434..189B}.}, possibly enhanced due to compression of the magnetic field, while the fainter emission from the rest of the PWN could be detected in deeper high-resolution X-ray observations. We should note that such a small head would not be resolved if J0437  were not so nearby. Among other nearby pulsars, Geminga ($d\sim 250$ pc) and PSR B1055--52 ($d\sim 350$ pc) show hints of X-ray emission up to a few arcseconds ahead of the pulsar \citep{2010ApJ...715...66P,2015ApJ...811...96P}. However, Geminga also shows three tails whose origin is unclear, while B1055--52 is likely moving away from us close to the line of sight, which leads to a different shape of the X-ray PWN image.

%
%
%

The X-ray luminosity of the detected extended emission, $L_{\rm 0.5-8\,keV} \approx3\times10^{28}$ erg\,s$^{-1}$, corresponds to the X-ray efficiency $\eta_{\rm 0.5-8\,keV}=L_{\rm 0.5-8\,keV}/\dot{E}=0.6\times10^{-5}$. Such an efficiency is consistent with those of PWNe of  supersonically moving pulsars \citep{2008ApJ...684..542K}, which supports our interpretation of the extended X-ray emission. 

Although the interpretation of the observed FUV emission as due to a collisionless shock in a partially ionized ISM is consistent with the data available, we cannot provide a direct proof of this interpretation until the emission spectrum (including spectral lines) is measured and compared with bow shock model predictions. Therefore, there still remains a possibility that the FUV emission might be due to synchrotron radiation of relativistic electrons leaked from the X-ray PWN region and trapped at the forward bow shock\footnote{Such kinetic processes are usually not taken into account by hydrodynamic simulations, such as those presented by \citet{2005A&A...434..189B}.}. To escape from the X-ray PWN and reach the bow shock, the electron should have a sufficiently large gyroradius, $r_g = 3.3\times 10^{15} (E_e/10\,{\rm TeV})(10\,\mu{\rm G}/B)$\,cm for the X-ray emitting electrons, at (and beyond) the PWN's outer boundary (cf.\ \citealt{2008A&A...490L...3B}). A fraction of the escaped electrons can be ``trapped'' in the bow shock vicinity  via multiple scattering on quasi-resonant magnetic fluctuations created by the current-driven instability (see, e.g., \citealt{2004MNRAS.353..550B,2011A&ARv..19...42B,2016RPPh...79d6901M}). Protons with energies of up to $\sim 0.1$--1 TeV, required for developing of this instability, could be accelerated at the moving bow shock. This scenario may result in a specific spectrum of the high energy electrons trapped in the bow shock vicinity  and producing the FUV synchrotron emission. Not only their energies should be above some minimal energy to escape from the X-ray PWN and reach the bow shock, but they should also be limited at the high energy end by the confinement condition at the bow shock. If this maximal energy is close to that of FUV-emitting electrons, it could explain the lack of X-ray emission at the bow shock. A sharp drop in the FUV surface brightness toward the X-ray PWN, such as the one seen in our SBC images, could  be due to a relatively low magnetic field in that region. Of course, this qualitative scenario remains hypothetical until it is supported by quantitative models.

To definitively determine the nature of the discovered FUV bow shock of PSR\,J0437$-$4715, its spatially resolved spectrum should be measured with a higher spectral resolution, in FUV as well as in NUV/optical wavelength ranges. We expect to see FUV shocks from other nearby pulsars, including those for which no H$\alpha$ bow shocks were detected. Observations of such FUV shocks, currently possible only with the {\sl HST}, would be very useful for studying the ISM properties, the interaction of pulsar winds with the ambient medium, and the mechanisms of particle acceleration.

\acknowledgements 
Support for programs GO\,12917 and GO\,10568 was provided by NASA through grants from the Space Telescope Science Institute, which is operated by the Association of Universities for Research in Astronomy, Inc., under NASA contract NAS 5-26555. A.M. Bykov acknowledges support from RSF grant 16-12-10225.  The authors are very grateful to Giovanni Morlino for the insightful discussion. A.M.B. and A.M.K. acknowledge Joint Supercomputer Center of Russian Academy of Science where some simulations were performed. We are thankful to the referee for the useful suggestions that helped to improve this paper.

\facilities{HST (ACS/SBC), CXO (ACIS)}


\end{document}